\documentclass[onecolumn]{IEEEtran}
\usepackage{graphics}
\usepackage{graphicx}
\usepackage{epsfig,amssymb}
\usepackage{amsmath}
\usepackage{mathrsfs}
\usepackage{color}
\usepackage{array}
\usepackage{amsfonts,booktabs}
\usepackage{lipsum,amsmath,multicol}
\usepackage{graphicx}
\usepackage{subfigure}
\usepackage{times}
\usepackage{pdfsync}
\usepackage{pgfplots}
\usepackage{pgfplotstable}
\usepackage[subnum]{cases}
\usepackage{filecontents}
\usepackage{multirow}
\usepackage{algorithm}
\usepackage{algpseudocode}
\usepackage{float}
\usepackage{setspace}
\newtheorem{theorem}{Theorem}
\newtheorem{lemma}{Lemma}
\newtheorem{definition}{Definition}
\newtheorem{corollary}{Corollary}

\newtheorem{remark}{Remark}

\newcommand{\beq}{\begin{equation}}
\newcommand{\eeq}{\end{equation}}
\newcommand{\beqa}{\begin{eqnarray}}
\newcommand{\eeqa}{\end{eqnarray}}

\newcommand{\paren}[1]{\left(#1\right)}
\newcommand{\sqparen}[1]{\left[#1\right]}
\newcommand{\brparen}[1]{\left\{#1\right\}}
\newcommand{\field}[1]{\ensuremath{\mathbb{#1}}}
\newcommand{\abs}[1]{\left|#1\right|} % absolute value
\newcommand{\N}{\ensuremath{\field{N}}} % natural numbers
\newcommand{\R}{\ensuremath{\field{R}}} % real numbers
\newcommand{\Z}{\ensuremath{\field{Z}}} % integers
\newcommand{\PR}[1]{\ensuremath{\mathsf{Pr}\left\{#1\right\}}} % probability with braces
\newcommand{\PRP}[1]{\ensuremath{\mathsf{Pr}\left(#1\right)}} %\probability with parentheses
\newcommand{\ES}[1]{\ensuremath{\mathsf{E}\left[#1 \right]}} %Expectation with square parentheses
\newcommand{\BO}[1]{\ensuremath{O\paren{#1}}}

\renewcommand{\vec}[1]{\ensuremath{\boldsymbol{#1}}}

\newcommand{\diag}[1]{\ensuremath{{\rm \bf diag}\paren{#1}}}

\newcommand{\logp}[1]{\ensuremath{\log\paren{#1}}}

\newcommand{\norm}[2]{\left\Vert#1\right\Vert_{#2}}

\newcommand{\ESI}[2]{\ensuremath{\mathsf{E}_{#1}\left[#2 \right]}} %Expectation with square parenthesesand index
\newcommand{\act}[3]{\ensuremath{#1^{#2}_{#3}}}%action of users
\newcommand{\CP}[2]{\ensuremath{\mathsf{Pr}\paren{\left.#1\right|#2}}}%conditional probability
\newcommand{\CD}[2]{\ensuremath{p_{#1}\paren{\vec{x}\left|#2\right.}}}%conditional density
\newcommand{\CDE}[2]{\ensuremath{\mathsf{h}\sqparen{#1\left|#2\right.}}}%conditional differntial entropy
\newcommand{\DE}[1]{\ensuremath{\mathsf{h}\sqparen{#1}}}%differential entropy  
\newcommand{\DSE}[1]{\ensuremath{\mathsf{H}\sqparen{#1}}}%discrete entropy  
\newcommand{\CDSE}[2]{\ensuremath{\mathsf{H}\sqparen{#1\left|#2\right.}}}%conditional discrete entropy
\newcommand{\MI}[2]{\ensuremath{\mathsf{I}\sqparen{#1;#2}}}%Mutual information 
\newcommand{\CMI}[3]{\ensuremath{\mathsf{I}\sqparen{#1;#2\left|#3\right.}}}%conditional mutual information
\newcommand{\KLD}[2]{\ensuremath{\mathsf{D}\sqparen{#1\left\|#2\right.}}}%Kulback libler distance 
\begin{document}

\title{Lower Bounds on the Complexity of  Solving Two Classes of Non-cooperative Games}
\author{Ehsan Nekouei, \IEEEmembership{Member, IEEE},  Girish N.~Nair, \IEEEmembership{Member, IEEE}, Tansu Alpcan, \IEEEmembership{Senior Member, IEEE}, Robin J.~Evans, \IEEEmembership{Life Fellow, IEEE}
\thanks{This work was supported by the Australian Research Council's Discovery Projects funding scheme (DP140100819). Department of Electrical and Electronic Engineering, The University of Melbourne, VIC 3010, Australia. E-mails: \{ehsan.nekouei,gnair,tansu.alpcan,robinje\}@unimelb.edu.au. This paper was presented in part at NecSys 2016 workshop, Tokyo, September 2016.  }}
\maketitle
\thispagestyle{empty}

\begin{abstract}
This paper studies the complexity of solving two classes of non-cooperative games in a distributed manner in which the players communicate with a set of system nodes over noisy communication channels.  The complexity of solving each game class is defined as the minimum number of iterations required  to find a Nash equilibrium (NE) of  any game in that class with  $\epsilon$ accuracy. First, we consider the class $\mathcal{G}$ of all $N$-player non-cooperative games with a continuous action space that admit at least one NE. Using information-theoretic inequalities, we derive a lower bound on the complexity of solving $\mathcal{G}$ that depends on the Kolmogorov $2\epsilon$-capacity of the constraint set and the total capacity of the communication channels. We also derive a lower bound on the complexity of solving  games in $\mathcal{G}$ which depends on the volume and surface area of the constraint set.  We next consider the class of all $N$-player non-cooperative games with at least one NE such that the players' utility functions satisfy a certain (differential) constraint. We derive lower bounds on the complexity of solving this game class under both Gaussian and non-Gaussian noise models. Our result in the non-Gaussian  case is derived by establishing a connection between the Kullback-Leibler distance and Fisher information.
%The complexity of  a subset of $\mathcal{G}$, and the norm of Jacobian of the pseudo-gradient vector, induced by utility functions of players, is more than $\gamma$ for all games in  $\mathcal{G}_\gamma$. Our results show that the complexity of solving the game class $\mathcal{G}_\gamma$ is lower bounded as $\Omega\paren{\frac{1}{\gamma^2\epsilon^2}}$ when the communication noise terms are Gaussian distributed. We also show that the same lower bound continues to hold on the complexity of solving the game class $\mathcal{G}_\gamma$ under noise-free  communication channels from the players to system nodes and non-Gaussian communication channels from system nodes to players. To establish this result, we show that the Kullback-Leibler distance between a non-Gaussian probability distribution function (PDF) and its shifted version can be written, up to an error term, as a monomial  which is quadratic in the shift parameter and linear in the Fisher information of the corresponding PDF with respect to the shift parameter.
\end{abstract}

\section{Introduction}

\subsection{Motivation}
Game theory offers a suite of analytical frameworks for investigating the interaction between rational decision-makers, hereafter called players. In the past decade, game theory has found diverse applications across engineering disciplines ranging from power control in wireless networks to modeling the behavior of travelers in a transport system.  The {\em Nash Equilibrium (NE)} is the fundamental solution concept for non-cooperative games, in which a number of players compete to maximize conflicting utility functions that are influenced by the action of others. At the NE, no player benefits from a unilateral deviation from its NE strategy. 

Finding the NE of a non-cooperative game is a fundamental research problem that lies at the heart of game theory literature. For non-cooperative games with continuous action spaces, various Nash seeking algorithms have been proposed in the literature, \emph{e.g.} see \cite{LB87}, \cite{LK11}.  In this paper, we investigate the intrinsic difficulty of finding a NE in such games. Using the notion of complexity from the convex optimization literature, and information-theoretic inequalities, we derive lower bounds on the minimum number of iterations required to find a NE within a desired accuracy, for
any  $N$-player, non-cooperative game in a given class.

\subsection{Related Work}

The book by \cite{NY83} pioneered the investigation of complexity in convex optimization problems. In this model, an algorithm sequentially queries an \emph{oracle} about the objective function of a convex optimization problem, and the oracle responds according to the queries and the objective function. 
They derive bounds on the minimum number of queries required to find the global optimizer of any function in a given function class. In \cite{AWPP09}, information-theoretic lower bounds were derived on the complexity of convex optimization problems with a stochastic first order oracle for the class of functions with a known Lipschitz constant. \textcolor{black}{In a stochastic first order oracle model, the algorithm receives randomized information about the objective function and its subgradient.} These results were extended to different function classes in \cite{AWPP12}. 

The paper \cite{RR11} considered a model in which the algorithm observes noisy versions of the oracle's response and established lower bounds on the complexity of convex optimization problems under first order as well as gradient-only oracles. In \cite{JNB12}, complexity lower bounds were obtained for convex optimization problems with a stochastic zero-order oracle. The paper \cite{DJWW12} studied  the complexity of convex optimization problems under a zero-order stochastic oracle in which the optimization algorithm submits two queries at each iteration and the oracle responds to both queries. These results were extended to the case in which the algorithm makes queries about multiple points at each iteration in \cite{DJWW15}. In \cite{SV15}, the complexity of convex optimization problems was studied under an erroneous oracle model wherein the oracle's responses to queries are subject to absolute/relative errors.  
\subsection{Contributions}
This paper studies the complexity of solving two classes of non-cooperative games in a distributed setting in which players communicate, not with an oracle, but with a set of utility system nodes (USNs) and constraint system nodes (CSNs) to obtain the required information for updating their actions. Each USN computes the utility-related information for a subset of players whereas a CSN evaluates a subset of constraint functions. The communication between players and system nodes is  subject to noise,
%regarding regarding its utility function and constraint set from its dedicated system node (SN). Each SN has access to the queries of all players which are coupled with its associated players  either through utility function or through constraints.
\emph{i.e.,} the system nodes will receive noisy versions of players' actions, and the players will receive noisy information from the system nodes.% We refer to the communication channels from players to system nodes as uplink channels and the communication channels from system nodes to players as downlink channels.

First, we consider the game class $\mathcal{G}$ which consists of all $N$-player non-cooperative games with a joint action space defined by $L$ convex constraints such that all the games in $\mathcal{G}$ admit at least one Nash equilibrium (NE). We derive lower bounds on the minimum number of iterations required to get within $\epsilon$ of a  NE of any game in $\mathcal{G}$ with confidence $1-\delta$ without imposing any particular structure on the computation model at USNs. Our results indicate that the complexity of solving the game class $\mathcal{G}$  is limited by the  Kolmogorov $2\epsilon$-capacity of the constraint set and the total capacity of communication channels from the USNs to the players. We also derive a lower bound on the complexity of solving the game class $\mathcal{G}$ in terms of the volume and surface area of the constraint set. We note that, in a precursor conference paper \cite{NANE16},  we have studied the complexity of solving the game class $\mathcal{G}$ under a slightly different setting than that in the current manuscript.

We next consider the game class $\mathcal{G}_\gamma$ which consists of all non-cooperative games with a joint action space defined by $L$ constraints such that $(i)$ all the games in $\mathcal{G}_\gamma$ admit at least one NE, $(ii)$ the norm of the Jacobian matrix of the \emph{pseudo-gradient} vector, induced by utility functions of players, is more than $\gamma$. We study the complexity of solving the game class $\mathcal{G}_\gamma$ under the partial-derivative computation model at USNs and various  noise models. Under the partial-derivative computation model,  each player receives a noisy version of the partial derivative of its utility function, with respect to its action,  in each iteration.

Our results show that the complexity of solving the game class $\mathcal{G}_\gamma$ up to $\epsilon$ accuracy is at least of order ${\frac{1}{\gamma^2\epsilon^2}}$, as $\epsilon$ tends to zero, with Gaussian  communication channels. We also consider a setting in which the channels from system nodes to players are non-Gaussian and the channels from players to system nodes are noiseless.  In the non-Gaussian setting, our results show that the complexity of solving the game class $\mathcal{G}_\gamma$  up to $\epsilon$ accuracy is at least of order ${\frac{1}{\gamma^2\epsilon^2}}$ as $\epsilon$ tends to zero. This result is established by deriving an asymptotic expansion for the Kullback-Leibler (KL) distance between a non-Gaussian probability distribution function (PDF) and its shifted version, under some mild assumptions on the non-Gaussian PDF. More precisely, it is shown that the KL distance between a PDF and its shifted version can be written, up an error term, as a monomial which is quadratic in the shift parameter and linear in the  Fisher information of the corresponding PDF with respect to the shift parameter.
 
This paper is organized as follows. Section \ref{Sec: SM} discusses our modeling assumptions and problem formulation. Section \ref{Sec: R&D} discusses our main results along with their interpretations. All the proofs are relegated to Section \ref{Sec: Derivations} to improve the readability of the paper. Section \ref{Sec: Conc} concludes the paper. 

\section{System Model}\label{Sec: SM}
\begin{figure*}[t]
\centering
\includegraphics[scale=0.7]{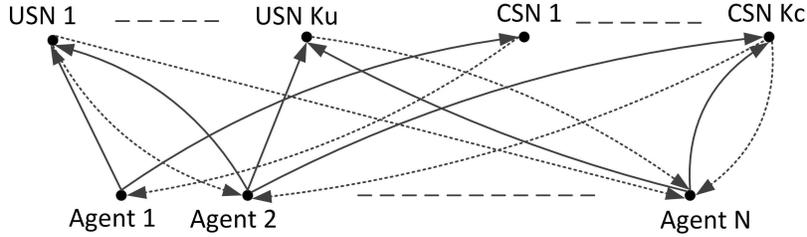}
\caption{\small A pictorial representation of the communication graph between  system nodes and players. Solid arrows denote the uplink channels and dashed arrows denote the downlink channels.  }
\label{F1}
\end{figure*}
\subsection{Game-theoretic Set-up}
Consider a non-cooperative game with $N$ players indexed over $\mathcal{N}=\left\{1,\cdots,N\right\}$. Let $x^i$ ($i\in\mathcal{N}$), and $\vec{x}=\left[x^1,\cdots,x^N\right]^\top$ denote  the action of the $i$th player and the collection of all players' actions, respectively. The utility function of the $i$th player is denoted by $u_i\paren{x^i,\vec{x}^{-i}}$ where $\vec{x}^{-i}$ is the vector of those other players' actions that affect the $i$th player's utility. The utility function of the $i$th player quantifies the desirability of any point in the action space for the $i$th player. The actions of players are limited by $L$ convex constraints denoted by $\vec{g}\paren{\vec{x}}\leq0$ where $\vec{g}\paren{\cdot}=\left[g_1\paren{\cdot},\cdots,g_L\paren{\cdot}\right]^\top$ is a mapping from $\R^N$ to $\R^L$. The set of constraint functions is indexed over $\mathcal{L}=\left\{p\in\N:1\leq p\leq L\right\}$. Let $\mathcal{S}$ denote the action space of players, \emph{i.e.,} 
\begin{eqnarray}
\mathcal{S}=\left\{\vec{x}\in\R^{N} s.t.\, \vec{g}\paren{\vec{x}}\leq 0\right\}.\nonumber
\end{eqnarray}
 We assume that $\mathcal{S}$ is a compact and convex subset of $\R^N$. 

In non-cooperative games, each player is interested in maximizing its own utility function, irrespective of other players. Since the maximizers of utility functions of players do not necessarily coincide with each other, a trade-off is required. In this paper, the Nash equilibrium is considered as the canonical solution concept of the non-cooperative game among players. Let  $\vec{x}_{\rm NE}\in\mathcal{S}$ be the NE of the game among players. Then, at the NE, no player has incentive to unilaterally deviate  its action from its NE strategy, \emph{i.e.,}
  \begin{eqnarray} 
\act{x}{i}{\rm NE}=\arg\underset{x^i\in\mathcal{S}\paren{\act{\vec{x}}{-i}{\rm NE,C}}}{\max} & u_i\paren{\act{x}{i}{},\act{\vec{x}}{-i}{\rm NE}}, \forall i\in\mathcal{N},\nonumber 
 \end{eqnarray}
\textcolor{black}{where $\vec{x}^{-i}_{\rm NE,C}$ is the collection of NE strategies of players which are coupled with the $i$th player through constraints, and  $\mathcal{S}\paren{\act{\vec{x}}{-i}{\rm NE,C}}$ is the set of possible actions of the $i$th player given $\act{\vec{x}}{-i}{\rm NE,C}$.} The vector of all utility functions is denoted by $U\paren{\vec{x}}=\left[u_1\paren{\vec{x}},\cdots,u_N\paren{\vec{x}}\right]^\top$. 

Let $\mathcal{F}$ denote the class of functions from $\R^N$ to $\R^N$ such that any $N$-player non-cooperative game with the constraint set $\mathcal{S}$ and utility function vector in $\mathcal{F}$ admits at least one NE. By the class of non-cooperative games $\mathcal{G}=\langle \mathcal{N}, \mathcal{S},\mathcal{F}\rangle$, we mean the set of all games with $N$ players, the action space $\mathcal{S}$, and the utility function vector in $\mathcal{F}$, \emph{i.e.,} $U\paren{\cdot}\in\mathcal{F}$.
 
%Let $G_{\rm u}=\left[G_{ij}\right]$ be an $N$-by-$N$ matrix whose $i,j$th entry is equal to one if the action of the $i$th player affects the utility function of the $j$th player, and is equal to zero otherwise. Similarly, let $G_{\rm c}=\left[G_{ij}\right]$ be an $N$-by-$L$ matrix whose $i,j$th entry is equal to one if the $j$th constraint is affected by the $i$th player's action, and is equal to zero otherwise. We refer to $G_{\rm u}$ and $G_{\rm c}$ as the utility adjacency and constraint adjacency matrices, respectively.
	\subsection{Communication Model}

In this paper, we consider a distributed Nash seeking set-up wherein, at each time-step, players  communicate with a set of utility system nodes (USNs) and constraint system nodes (CSNs) to obtain the required utility/constraint related information for updating their actions. A USN computes utility-related information for a set of players, \emph{e.g.,} the utility functions of players or their partial derivatives.  A CSN evaluates a subset of constraints based on the received actions of players. Each utility function or constraint is evaluated at only one USN or CSN, respectively. The number of USNs and CSNs are denoted by $K_{\rm u}$ and $K_{\rm c}$, respectively, with $K_{\rm u}\leq N$ and $K_{\rm c}\leq L$. We use $\text{USN}_l$ ($l\in\left\{1,\dots,K_{\rm u}\right\}$) and  $\text{CSN}_n$ ($n\in\left\{1,\dots,K_{\rm c}\right\}$) to refer to the $l$th USN and $n$th CSN, respectively. 

At each time-step, player $i$ transmits its action to $\text{USN}_l$ if its action affects at least a utility function evaluated by $\text{USN}_l$. The set of players which transmit their actions to $\text{USN}_l$ is denoted by $\mathcal{N}_{{\rm usn}_l}$.  We use the mapping $\pi\paren{\cdot}$, from $\left\{1,\cdots,N\right\}$ to $\left\{1,\dots,K_{\rm u}\right\}$, to indicate the USN which computes the utility-related information for a given player, \emph{i.e.,} $\pi\paren{i}=l$ if  $\text{USN}_l$ computes the utility-related information for the $i$th player. Thus, the utility-related information for the $i$th player is computed by $\text{USN}_{\pi\paren{i}}$.

Similarly, at each time-step, player $i$ transmits its action to $\text{CSN}_n$ if its action affects at least one constraint function evaluated by $\text{CSN}_n$. The set of players which transmit their actions to $\text{CSN}_n$ is represented by $\mathcal{N}_{{\rm csn}_n}$. We use the mapping $\phi\paren{\cdot}$, from  $\left\{1,\cdots,L\right\}$ to $\left\{1,\dots,K_{\rm c}\right\}$, to indicate the CSN which evaluates a given constraint function, \emph{i.e.,} $\phi\paren{p}=n$ if $\text{CSN}_n$ evaluates the $p$th constraint function. Hence, the $p$th constraint function is evaluated by $\text{CSN}_{\phi\paren{p}}$. The set of constraint functions which are affected by the $i$th player's action are denoted by $\mathcal{L}_i$. 

%{\bf only if, guarantee that system nodes have the required information for their computations. } Through this paper,  Similarly, 
%The following notations are used through this paper. We use $i\in\mathcal{N}$ to refer to an player, $l$ to refer to a USN, and $l_i$ to denote the USN which provides utility-related information for the $i$th player. We reserve $p\in\mathcal{L}$ to denote a constraint, $n$ to refer to a CSN, and $n_p$ to represent the CSN which evaluates the $p$th constraint. The time index is denoted by $k\in\N$, and the set of constraints which depend on the action of the $i$th are denoted by $\mathcal{L}_i\subset\mathcal{L}$. The notations $\mathcal{N}_{\rm usn}^{l}\subset\mathcal{N}$ and $\mathcal{N}^n_{\rm csn}\subset\mathcal{N}$ denote the set of players which transmit their actions to the $l$th USN and the set of players which transmit their actions to the $n$th CSN, respectively. Note that if $i\in\mathcal{N}^l_{\rm usn}$ ($i\in\mathcal{N}^n_{\rm csn}$), then, the $i$th player's action affects at least a utility function (constraint) evaluated by the $l$th USN ($n$th CSN), thus, the $l$th USN ($n$th CSN) requires the $i$th player's action for its computations. %are evaluated by the $n$th CSN, and the set of players which their actions are required by the $n$th CSN to evaluate the constraints in $\mathcal{L}^{\rm csn}_{n}$, respectively. The system nodes are designed such that we have $\mathcal{N}=\cup_l\mathcal{N}^{\rm usn}_{l}$ and $\mathcal{L}=\cup_n\mathcal{L}^{\rm csn}_{n}$. 

 The communication topology between players and system nodes is given by a bipartite digraph in which the players and the system nodes form two disjoint sets of vertices. There exists a directed edge, in the communication graph, from the $i$th player to $\text{USN}_l$ if $i\in\mathcal{N}_{{\rm usn}_l}$. Also, there exists a directed edge from $\text{USN}_{\pi\paren{i}}$ to the $i$th player for all $i\in\mathcal{N}$. Furthermore, there exist a directed edge from the $i$th player to $\text{CSN}_n$, and a directed edge from $\text{CSN}_n$ to the $i$th player if $i\in\mathcal{N}_{{\rm csn}_n}$. We refer to communication channels from  players to system nodes as uplink channels and the communications channels between system nodes and players as downlink channels. %{\bf a figure showing the uplink and downlink channels?} 
Fig. \ref{F1} shows a pictorial representation of the communication topology between system nodes and  players.

Players communicate with system nodes using frequency division multiplexing (FDM) or time division multiplexing (TDM) schemes, \emph{i.e.,} each player broadcasts its action to its neighboring system nodes in the communication graph using a dedicated time or frequency band. Similarly, system nodes communicate with players via FDM or TDM communication schemes. The communication between players and system nodes is performed over noisy communication channels, \emph{i.e.,} players receive noisy information from system nodes, and system nodes receive noisy versions of players' actions. This will be made more explicit in the next subsection.

 % , and transmits the constraint values to the players which their actions affect the constraints. Each player broadcasts its action to its neighboring SNs, and each SN broadcasts  utility/constraint information to its neighboring players. The $j$th SN is responsible for evaluating a subset of the constraints, denoted by $\mathcal{L}_j$, and/or providing utility-related information for a subset of players, denoted by $\mathcal{N}_j$, such that $\mathcal{N}=\cup_j\mathcal{N}_j$ and $\mathcal{L}=\cup_j\mathcal{L}_j$. Let $D_{\mathcal{N}}=\left[d_{ij}\right]$ denote an $N$-by-$N$ matrix whose $\left(i,j\right)$th entry is equal to 1 if the action of the $i$th player appears in the utility function of the $j$th player, and is  equal to zero otherwise. Also, let $G_{\mathcal{L}}=\left[g_{ij}\right]$ be an $N$-by-$l$ matrix whose $\left(i,j\right)$th entry is equal to 1 if the action of the $i$th player affects the $j$th constraint, and is equal to zero otherwise. There exists a directed edge, in the communication graph, from the $i$th player  to the $j$th SN if the $j$th  SN requires the $i$th player's action for its computations, \emph{i.e.,} if $d_{ip}=1$ for some $p\in\mathcal{N}_j$ or $g_{ip}=1$ for some $p\in\mathcal{L}_j$. Similarly, there exists a directed edge from $j$th SN  to the $i$th player if the $j$th SN provides utility-related information to the $i$th player or evaluates a constraint which involves the $i$th player's action, \emph{i.e.,}  $i\in \mathcal{N}_j$ or if  $g_{ip}=1$ for $p\in\mathcal{L}_j$.  
\begin{figure*}
\begin{eqnarray}\label{Eq: Complexity_Def}
T^\star_{\epsilon,\delta}\paren{\mathcal{G},\mathcal{O}}=\inf\brparen{T\in\N: \exists \mathcal{A}\quad  s.t. \sup_{U\paren{\cdot}\in \mathcal{F}}\inf_i\PRP{\norm{\vec{x}_{{\rm NE}_i, U\paren{\cdot}}-\mathcal{A}_{T+1}\paren{{X}_{1:T},{\hat{Y}_{1:T}},\hat{Z}_{1:T}}}{}\geq \epsilon}\leq \delta}.
\end{eqnarray}
\hrule
\end{figure*}
\subsection{Nash Seeking Algorithms}
\subsubsection{The Update Rule}
In this paper, we consider a general structure for the Nash seeking algorithms which allows each player's action to be updated using the past actions of that player as well as the past received utility/constraint related information by that player.
Let $\mathcal{A}$ be such a Nash seeking algorithm. Then, under $\mathcal{A}$, the $i$th player's action at time $k$, \emph{i.e.,} $x^i_k$, is updated according to the update rule 
\begin{align}
x^i_k=\mathcal{A}^i_k\paren{X^i_{1:k-1},\hat{Y}^i_{1:k-1},\hat{Z}^i_{1:k-1}},\nonumber 
\end{align}
where $X^i_{1:k-1}$ is the history of the $i$th player's actions from time $1$ to $k-1$, $\hat{Y}^i_{1:k-1}$ denotes the sequence of received utility-related information by the $i$th player from time $1$ to $k-1$, and $\hat{Z}^i_{1:k-1}$ denotes the sequence of received constraint-related information by the $i$th player from time $1$ to $k-1$. Here, $\mathcal{A}^i_k\paren{\cdot,\cdot,\cdot}$ is a mapping from $\R^{k-1}\times\R^{k-1}\times\R^{\paren{k-1}\abs{\mathcal{L}_i}}$ to $\R$. Note that   $X^i_{1:k-1}$, $\hat{Y}^i_{1:k-1}$ and $\hat{Z}^i_{1:k-1}$ can be written as 
\begin{align}
X^i_{1:k-1}&=\left\{x^i_t\right\}_{t=1}^{k-1},\nonumber\\
\hat{Y}^i_{1:k-1}&=\left\{\hat{y}^i_t\right\}_{t=1}^{k-1},\nonumber\\
\hat{Z}^i_{1:k-1}&=\left\{\hat{z}^{i,p}_t,p\in\mathcal{L}_i\right\}_{t=1}^{k-1},\nonumber
\end{align}
, respectively, where  $x^i_t$ is the action of the $i$th player at time $t$, $\hat{y}^i_t$ denotes the received utility-related information by the $i$th player at time $t$ and  $\hat{z}^{i,p}_t$ denotes the received information regarding the $p$th constraint by the $i$th player at time $t$. 

The $k$th step of the algorithm $\mathcal{A}$ is denoted by 
	\begin{align}
&\mathcal{A}_k\!\!\paren{X_{1:k-1},\hat{Y}_{1:k-1},\hat{Z}_{1:k-1}}\!\!=\!\!\left\{\!\mathcal{A}^i_k\!\!\paren{X^i_{1:k-1},\hat{Y}^i_{1:k-1},\hat{Z}^i_{1:k-1}\!\!}\right\}_{i},\nonumber
	\end{align}
	where
	\begin{align}
	X_{1:k-1}&=\left\{x^i_t:i\in\mathcal{N}\right\}_{t=1}^{k-1},\nonumber\\
	\hat{Y}_{1:k-1}&=\left\{\hat{y}^i_{t}:i\in\mathcal{N}\right\}_{t=1}^{k-1},\nonumber\\
	\hat{Z}_{1:k-1}&=\left\{\hat{z}^{i,p}_t:i\in\mathcal{N},p\in\mathcal{L}_i\right\}_{t=1}^{k-1}.\nonumber
	\end{align}
	%	$X_{1:k-1}=\left\{x^i_t:i\in\mathcal{N}\right\}_{t=1}^{k-1}$ is the history of all players' actions from time $1$ to $k-1$, $\hat{Y}_{1:k-1}=\left\{\hat{y}^i_{t}:i\in\mathcal{N}\right\}_{t=1}^{k-1}$ is the history of all received utility-related information by players from time $1$ to $k-1$, and $\hat{Z}_{1:k-1}=\left\{\hat{z}^i_{t,p}:i\in\mathcal{N},p\in\mathcal{L}_i\right\}_{t=1}^{k-1}$ is the history of all received constraint-related information by players from time $1$ to $k-1$. 
	
	We refer to 
	\begin{align} 
	\mathcal{A}=\left\{\mathcal{A}_k\paren{X_{1:k-1},\hat{Y}_{1:k-1},\hat{Z}_{1:k-1}}\right\}_{k},\nonumber
	\end{align}
	as the Nash seeking algorithm $\mathcal{A}$. 
	
%In what follows, we use $ W^{i}_{k,{\rm usn}_l}$ and $W^{i}_{k,{\rm csn}_n}$ to represent the additive noises associated with the transmission of $x^i_k$, \emph{i.e.,} the $i$th player's action at time $k$, to $\text{USN}_l$ and $\text{CSN}_n$, respectively. We also use $W_{k,i}$ and $W^{p}_{k,i}$ ($p\in\mathcal{L}_i$) to represent the additive communication noises associated with the transmission of utility-related information (by $\text{USN}_{\pi\paren{i}}$) and the value of the $p$th constraint (by $\text{CSN}_{\phi\paren{p}}$) to the $i$th player at time $k$, respectively.  
 
\subsubsection{Communication and Computation At USNs}	
  The received action of the $i$th player by $\text{USN}_l$ at time $k$, \emph{i.e.,} $\hat{x}^i_{k,{\rm usn}_l}$, can be written as 
	\begin{align}
	\hat{x}^{i}_{k,{\rm usn}_l}=x^i_k+W^{i}_{k,{\rm usn}_l}, \nonumber
	\end{align}
		where $W^{i}_{k,{\rm usn}_l}$	is the noise in the uplink channel from the $i$th player to $\text{USN}_l$.  	Let $\hat{X}_{1:k}^{{\rm usn}_l}=\left\{\hat{x}^{i }_{t,{\rm usn}_l}:i\in \mathcal{N}_{{\rm usn}_l}\right\}_{t=1}^k$ denote the history of actions received by $\text{USN}_l$ from time 1 to time $k$. At time $k$, $\text{USN}_{l}$  computes $y_{k}^i$, \emph{i.e.,} the utility-related information for player $i$ at time $k$, for all $i$ such that $\pi\paren{i}=l$. 
		
		In this paper, we study the complexity of solving non-cooperative games under two computation models at USNs. We first consider a general computation model in which $y^i_k$ is allowed to be any arbitrary function of $u_i\paren{\cdot}$ and the information available at $\text{USN}_{l}$ from time $1$ to $k$, \emph{i.e.,} 
	\begin{eqnarray}\label{Eq: Gen-Comp-Mod}
	y_{k}^i=\mathcal{O}_{k,i}\paren{\hat{X}_{1:k}^{{\rm usn}_{l}},u_i\paren{\cdot}} \quad \forall i: \pi\paren{i}=l.
	\end{eqnarray}
	where $\mathcal{O}_{k,i}\paren{\cdot,\cdot}$ is a functional.  %Then, the history of the received utility-related information by the $i$th player from time 1 to $k$ can be written as $\hat{Y}^i_{1:k}=\left\{\hat{y}^i_{t}\right\}_{t=1}^k$. 
%In \eqref{Eq: Gen-Comp-Mod}, we allow  
This formulation allows us to capture the complexity of solving the game class $\mathcal{G}$ under a general class of computation models at USNs in Theorem \ref{Theo: DownCap}.  We refer to $\mathcal{O}=\left\{\mathcal{O}_{k,i}\paren{\hat{X}^{{\rm usn}_{\pi\paren{i}}}_{1:k},u_i\paren{\cdot}}\right\}_{k,i}$ as the general computation model at USNs. 
	
	 We also study the complexity of solving non-cooperative games under the partial-derivative computation model in which  $\text{USN}_{l}$ at time $k$ evaluates the partial derivative of the utility function of the $i$th player with respect to its action, \emph{i.e.}, $y^i_k=\left.\frac{\partial }{\partial \paren{x^i}}u_i\paren{x^i,\vec{x}^{-i}}\right|_{\hat{X}^{{\rm usn}_{l}}_{k}}$ for all $i$ with $\pi\paren{i}=l$. We refer to the partial-derivative computational model for USNs as 
	\begin{align}\label{Eq: Der-only}
	\mathcal{O}^1=\left\{\mathcal{O}^1_{k,i}\paren{\hat{X}^{{\rm usn}_{\pi\paren{i}}}_{k},u_i\paren{\cdot}}\right\}_{k,i}
	\end{align}
	where $\hat{X}^{{\rm usn}_l}_{k}=\left\{\hat{x}^{i}_{k,{\rm usn}_l}:i\in \mathcal{N}_{{\rm usn}_l}\right\}$  denotes the set of actions received by $\text{USN}_l$ at time $k$ and 
\begin{align}
\mathcal{O}^1_{k,i}\paren{\hat{X}^{{\rm usn}_{\pi\paren{i}}}_{k},u_i\paren{\cdot}}=\left.\frac{\partial }{\partial \paren{x^i}}u_i\paren{x^i,\vec{x}^{-i}}\right|_{\vec{x}=\hat{X}^{{\rm usn}_{\pi\paren{i}}}_{k}}.\nonumber
\end{align}
	
	 Then, $\text{USN}_{l}$ transmits $y_k^i$ to the $i$th player for all $i$ with $\pi\paren{i}=l$. %For example, $y_{k}^i$ can be equal to $y_{k}^i=\left.\frac{\partial^n }{\partial \paren{x^i}^n}u_i\paren{x^i,\vec{x}^{-i}}\right|_{\hat{X}_{k}^{{\rm usn}_{\pi\paren{i}} }}$.
 The received utility-related information by the $i$th player at time $k$ can be written as 
\begin{align} 
\hat{y}^i_k=y^i_{k}+V^{i}_{k},\nonumber
\end{align}
 where $V^{i}_{k}$ is the noise in the downlink channel from the $\text{USN}_{\pi\paren{i}}$ to the $i$th player.
\subsubsection{Communication and Computation At CSNs}
The received action of the $i$th player by $\text{CSN}_n$ at time $k$, \emph{i.e,} $\hat{x}^{i}_{k,{\rm csn}_n}$, can be written as 
\begin{align}
\hat{x}^{i}_{k,{\rm csn}_n}=x^i_k+W^{i}_{k,{\rm csn}_n},\nonumber
\end{align}
where $W^{i }_{k,{\rm csn}_n}$ is the noise in the uplink channel from the $i$th player to $\text{CSN}_n$. The collection of received actions at time $k$ by the $\text{CSN}_n$ is denoted by $\hat{X}^{{\rm csn}_n}_k=\left\{\hat{x}^{i}_{k,{\rm csn}_n}:i\in\mathcal{N}_{{\rm csn}_{n}}\right\}$. At time $k$, $\text{CSN}_{n}$ evaluates its associated constraint functions using the received actions at time $k$, \emph{i.e.,}
\begin{align}
z^p_{k}=g_p\paren{\hat{X}_k^{{\rm csn}_{n}}}, \quad \forall p: \phi\paren{p}=n\nonumber
\end{align} 
Finally, $\text{CSN}_{n}$ broadcasts $z^p_{k}$ to the players which their actions affect $g_p\paren{\cdot}$. % via dedicated \emph{memoryless} channels. The received information by the $i$th player at time $k$ are denoted by $\hat{y}^i_k$ and $\hat{Z}^i_k$.% Fig. \ref{F1} shows the interaction between the PN and players at time $k$.
	If the action of the $i$th player affects the $p$th constraint, the $i$th player will  receive 
	\begin{align} 
	\hat{z}^{i,p}_{k}=z^p_{k}+V^{i,p}_{k},\nonumber
	\end{align}
	at time $k$ where $V^{i,p}_{k}$ is the noise in the downlink channel from $\text{CSN}_{\phi\paren{p}}$ to the player $i$. 
	\begin{remark}
	Although, we assume that the $\text{CSN}_{n}$ at time $k$ transmits $g_p\paren{\hat{X}_k^{{\rm csn}_{n}}}$ to the $i$th player (if $p\in\mathcal{L}_i$), our results continue to hold when other computation models are implemented at the CSNs, \emph{e.g.,} when the $\text{CSN}_{n}$ at time $k$ transmits $\left.\frac{\partial}{\partial x^i}g_p\paren{\vec{x}}\right|_{\vec{x}=\hat{X}_k^{{\rm csn}_{n}}}$ to the $i$th player.  
	\end{remark}
	 \begin{table*}[htp]
\caption{Table of the main variables }
  \begin{tabular}{ll}
		Variable &Description\\
    \hline \hline
$\mathcal{N}$& The set of players\\
$\mathcal{L}$& The set of constraints\\
%$p\in\mathcal{L}$& A constraint\\
%$i\in\mathcal{N}$&An player\\
$\mathcal{L}_i$&The set of constraints which are affected by the $i$th player action \\
%$l$& A utility system node (USN)\\
$\text{USN}_{\pi\paren{i}}$&The USN which computes utility-related information for the $i$th player\\
%$n$& A constraint system node (CSN)\\
$\text{CSN}_{\phi\paren{p}}$& The CSN which evaluates the $p$th constraint\\
$\mathcal{N}_{{\rm usn}_l}$&The set of players which transmit their actions to $\text{USN}_l$\\
$\mathcal{N}_{{\rm csn}_n}$&The set of players which transmit their actions to $\text{CSN}_n$\\
%$k$& The time index\\
$x^i_k$& Action of the $i$th player at time $k$\\
$X^i_{1:k}$& Actions of the $i$th player from time $1$ to $k$\\
%$X_k$& Actions of all the players at time $k$\\
$X_{1:k}$& Actions of all the players from time $1$ to $k$\\
$\hat{x}^{i}_{k,{\rm usn}_l}$& The received action of the $i$th player by $\text{USN}_l$ at time $k$\\
$\hat{X}^{{\rm usn}_l}_{k}$& The collection of received actions by $\text{USN}_l$ at time $k$\\
$\hat{X}^{{\rm usn}_l}_{1:k}$& The collection of received actions by $\text{USN}_l$ from time 1 to $k$\\
$y^i_k$&The utility-related information computed by $\text{USN}_{\pi\paren{i}}$ for the $i$th player\\
$\hat{y}^i_{k}$& The received utility-related information by the $i$th player\\
$\hat{Y}^i_{1:k}$& The history of received utility-related information by the $i$th players from time $1$ to $k$\\ 
$\hat{Y}_{1:k}$& The history of received utility-related information by all the players from time $1$ to $k$\\
$\hat{x}^{i}_{k,{\rm csn}_n}$& The received action of the $i$th player by $\text{CSN}_n$ at time $k$\\
$\hat{X}^{{\rm csn}_n}_{k}$& The collection of received actions by $\text{CSN}_n$ at time $k$\\
$z^p_k$& The value of the $p$th constraint at time $k$ evaluated by $\text{CSN}_{\phi\paren{p}}$\\
$\hat{z}^{i,p}_{k}$& The received value of the $p$th constraint at time $k$ by the $i$th player\\
$\hat{Z}^{i}_{1:k}$& The history of received constraint-related information by the $i$th players from time $1$ to $k$\\ 
$\hat{Z}_{1:k}$& The history of received constraint-related information by all the players from time $1$ to $k$\\
$W^{i}_{k,{\rm usn}_l}$& The additive noise in the uplink channel from the $i$th player to $\text{USN}_l$ at time $k$ ($i\in\mathcal{N}_{{\rm usn}_l}$)\\
$W^{i}_{k,{\rm csn}_n}$& The additive noise in the uplink channel from the $i$th player to $\text{CSN}_n$ at time $k$ ($i\in\mathcal{N}_{{\rm csn}_n}$)\\
$V^i_k$& The additive noise in the downlink channel from $\text{USN}_{\pi\paren{i}}$ to the $i$th player at time $k$\\
$V^{i,p}_k$& The additive noise in the downlink channel which transmits $z^p_k$ to the $i$th player at time $k$ ($p\in\mathcal{L}_i$)\\
	\hline
	\end{tabular}
  \label{Table: para}

\end{table*}
\subsection{The Complexity Criterion}
% The main body of game theory literature studies the performance of a particular algorithm, \emph{e.g.,} gradient, best response algorithms, etc, in finding the NE of a game when the objective functions of players pertain desirable properties, \emph{e.g.,} continuity, differentiability, etc. In contrast, in this paper, we do not restrict our analysis to a particular algorithm. Here, given a {\bf PN computation model}, we study the time performance of the fastest algorithm which solves all the games in a class of games. More precisely, we obtain lower bounds on the smallest number of time steps in which one can solve every game in a class of games using one algorithm with a given accuracy level and a PN computation model.  

Consider the class of games $\mathcal{G}$ and and the computation model $\mathcal{O}$. Then, the $\paren{\epsilon,\delta}$-complexity of solving the class of games $\mathcal{G}$ with the computation model $\mathcal{O}$, denoted by $T^\star_{\epsilon,\delta}\paren{\mathcal{G},\mathcal{O}}$, is defined in \eqref{Eq: Complexity_Def} where $\vec{x}_{{\rm NE}_i, U\paren{\cdot}}$ is a NE of the non-cooperative game with the utility function vector given by $U\paren{\vec{\cdot}}\in\mathcal{F}$. According to \eqref{Eq: Complexity_Def}, the $\paren{\epsilon,\delta}$-complexity of solving the class of games $\mathcal{G}$ with the computation model $\mathcal{O}$ is defined as the smallest positive integer $T$ for which there exists an algorithm $\mathcal{A}$ such that, for any game in $\mathcal{G}$, the probability of $\epsilon$ deviation of the algorithm's output at time $T+1$ from at least a  NE of the game is less than $\delta$.  Note that \eqref{Eq: Complexity_Def} assigns a positive integer to any class of games. For a given pair of $\paren{\mathcal{G},\mathcal{O}}$, a small value of $T^\star_{\epsilon,\delta}\paren{\mathcal{G},\mathcal{O}}$ indicates that the class of games $\mathcal{G}$ with the computation model $\mathcal{O}$ can be solved faster compared to a large value of $T^\star_{\epsilon,\delta}\paren{\mathcal{G},\mathcal{O}}$. The complexity of solving the game class $\mathcal{G}$ under the computation model $\mathcal{O}^1$ can be defined in a similar way.

\begin{remark}\label{Remark: NE}
The $\epsilon$-Nash equilibrium ($\epsilon$-NE) is a closely related solution concept to the NE which is defined as the point such that no play can gain more than $\epsilon$ by unilaterally deviating its strategy from its $\epsilon$-NE strategy. However, an $\epsilon$-NE is not always close to a NE  since game-theoretic problems are not necessarily convex problems and a NE is not necessarily the maximizer/minimizer of utility functions of all players \cite{LS08}. Hence, we do not consider $\epsilon$-NE as a solution concept in this  paper. %Since the current paper investigates the complexity of finding NE, we   is , in this paper, we consider the NE as the solution concept  %However, for the continuous, and strongly  convex function $f\paren{\vec{x}}$, the condition $f\paren{\vec{x}_{T+1}}-f^\star\leq \epsilon$ implies that $\vec{x}_{T+1}$ is close to the unique minimizer of $f\paren{\vec{x}}$. Thus, in this paper, we do not the $\epsilon$-NE as our solution concept.
\end{remark}
\subsection{Modeling Assumptions}\label{Subsec: Assum}
 	In this paper, we impose the following assumptions on the Nash seeking algorithms and the noise terms in the uplink/downlink communication channels: 
	\begin{enumerate}
%	\item The computation model $\mathcal{O}$ is not too informative. For example, the output of the map $\mathcal{O}$ cannot be the NE of the game among players. 
	%\item The players are aware of the computation model, \emph{i.e.,} the map $\mathcal{O}$ is known by players.
	\item $X_1$ is specified by the algorithm $\mathcal{A}$, and the algorithm $\mathcal{A}$ uses the same value of  $X_1$ for solving any game.  
	%\item {\bf remove} The communication channels are memoryless. 
	\item $\left\{W^{i}_{k,{\rm usn}_l},i\in\mathcal{N}_{{\rm usn}_l}\right\}_k$ is a collection of zero mean,  independent and identically distributed (i.i.d.) random variables with variance $\sigma^2_{{\rm usn}_l}>0$ for all $1\leq l\leq K_{\rm u}$. 
		\item $\left\{W^{i}_{k,{\rm csn}_n},i\in\mathcal{N}_{{\rm csn}_n}\right\}_k$ is a collection of zero mean, i.i.d. random variables  with variance $\sigma^2_{{\rm csn}_n}>0$ for all $1\leq n\leq K_{\rm c}$.
		\item $\left\{V^i_k,V^{i,p}_k,p\in\mathcal{L}_i\right\}_k$ is a collection of i.i.d. random variables with zero mean and variance $\sigma^2_i>0$ for all $i\in\mathcal{N}$. 
				%\item $\left\{W^i_{k,n_p},n_p\in\mathcal{L}_i\right\}_k$ is a collection of i.i.d. random variable with zero mean and variance  for all $i\in\mathcal{N}$. 	
				\item		All the uplink/downlink noise terms are jointly independent. %The collection of random variables 
				%\item The Lebesgue measure on $\R$ and the  probability measure of each noise term are mutually absolutely continuous.\footnote{Let $\lambda\paren{x}$ and $\mu\paren{x}$ denote two measures on $\R$. Then, $\lambda\paren{x}$ and $\mu\paren{x}$ are mutually absolutely continuous if $\lambda\paren{x}\ll\mu\paren{x}$ and $\mu\paren{x}\ll\lambda\paren{x}$ where $\lambda\paren{x}\ll\mu\paren{x}$ denotes the absolute continuity of $\lambda\paren{x}$ with respect to $\mu\paren{x}$.}
	%\begin{eqnarray}
	%&\hspace{-3cm}\left\{W^{i,l}_{k,{\rm usn}},W^{i^\prime,n}_{k,{\rm csn}},W^j_{k,l_j},W^{j^\prime}_{k,n_p}, 1\leq l\leq K_{\rm u},\right.\nonumber\\
	%&\left.  1\leq n\leq K_{\rm c}, i\in\mathcal{N}^l_{\rm usn},i^\prime\in \mathcal{N}^p_{\rm csn},j\in\mathcal{N},p\in\mathcal{L}^{j^\prime}\right\}_k
	%\end{eqnarray}
	%are jointly independent. 
	%\item The game $\mathcal{G}$ admits a unique NE for any $U\paren{\cdot}\in\mathcal{F}$ (we might be able to remove this assumption)
%\item The utility functions of players are continuously differentiable functions. 
		\end{enumerate}

%The efficiency of the ES algorithm $\mathcal{A}$ is defined as the smallest number time steps required to find the NE of any game $\mathcal{G}$ with $U\paren{\vec{x}}\in\mathcal{F}$, up to an error $\epsilon$, with high probability. More precisely, the efficiency of the algorithm $\mathcal{A}$, denoted by $T^\star_\mathcal{A}$, is defined as    
%\subsection{Table of Notations}
%Table \ref{Table: para} summarizes our main notations. 

%Let $X_k$ and $\hat{Y}_k$ denote the collection of submitted action of all players at time $k$ and the collection of  received action of all player by PN at time $k$. We use $y^i_k$ and $\hat{x}^i_k$ denote the response of PN to the $i$th player at time $k$, and the received action of the $i$th player by PN at time $k$. For notational convenience, we use $\hat{X}_k$, $Y_k$ to denote the collections of the received actions all players by PN at time $k$ and the vector of PN responses to players at time $k$. The collections of received actions of the $i$th player by PN from time 1 to $k$ is denoted by $\hat{X}^i_{1:k}$, and the vector of response of PN to the $i$th player from time 1 to $k$ is denoted by $Y^i_{1:k}$. Finally, the collection of all received actions of players from time 1 to $k$ is denoted by $\hat{X}_{1:k}$ and the collection of the responses of PN to all players from time 1 to $k$ is denoted by $Y_{1:k}$. Different elements of $\mathcal{F}$ are differentiated by  different subscripts, \emph{e.g.,} $U_i\paren{\vec{x}}$ and $U_j\paren{\vec{x}}$.
 \subsection{Organization of The Paper and Notations}
The rest of this paper is organized as follows. Section \ref{Sec: R&D} states our main results on the complexity of solving two classes of non-cooperative games. Section \ref{Sec: Derivations} presents the derivation of our results, and Section \ref{Sec: Conc} concludes the paper. 

In the rest of this paper, we use the following notations from asymptotic analysis literature. For two positive functions $f\paren{x}$ and $g\paren{x}$, we say $f\paren{x}=\Omega\paren{g\paren{x}}$ if $\lim\inf_{x\downarrow 0}\frac{f\paren{x}}{g\paren{x}}> 0$. We also say $f\paren{x}=\Theta\paren{g\paren{x}}$ if $\lim\inf_{x\downarrow 0}\frac{f\paren{x}}{g\paren{x}}> 0$ and $\lim\sup_{x\downarrow 0}\frac{f\paren{x}}{g\paren{x}}<\infty$. Our main notations are summarized in Table \ref{Table: para}.
% {\bf define $\Omega\paren{\cdot}$ and $\Theta\paren{\cdot}$}
\section{Results and Discussions}\label{Sec: R&D}
In this section, we establish various lower bounds on the complexity of solving two game classes under different assumptions on the distribution of uplink/downlink noise terms and different computation models at USNs. In Subsection \ref{Sec: General}, we derive two lower bounds on the complexity of solving the game class $\mathcal{G}$ under the general computation model at USNs without assuming any particular distribution for the uplink/downlink noise terms. In Subsection \ref{Subsec: Gauss}, we establish a lower bound on the complexity of solving a subclass of $\mathcal{G}$, denoted by $\mathcal{G}_\gamma$, under Gaussian uplink/downlink channels and the partial-derivative computation model. Subsection \ref{Subsec: Gauss} presents a lower bound on the complexity of solving the game class $\mathcal{G}_\gamma$ under noiseless uplink channels, non-Gaussian downlink channels, and the partial-derivative computation model. Subsection \ref{Sec: Gen-up-down} discusses  the complexity of solving the game class $\mathcal{G}_\gamma$ under the partial-derivative computation model when both uplink and downlink channels are non-Gaussian distributed.
\subsection{General Computation model at USNs and General Uplink/Downlink Channels}\label{Sec: General}
In this subsection, we study the computational complexity of solving the game class $\mathcal{G}$ without imposing any particular structure on the computation model at the USNs, or imposing any specific probability distribution on the noise in the uplink/downlink channels. To this end, we first give the definition of the total capacity of downlink channels, the notion of $2\epsilon$-distinguishable subsets of $\mathcal{S}$, and the Kolmogorov capacity of $\mathcal{S}$.

The total capacity of downlink channels \textcolor{black}{from USNs to players} is defined as $$C_{\rm down}=\max_{p_{Y}\paren{\vec{y}},\ES{\norm{Y}{}^2}\leq \alpha}\MI{y^1,\cdots,y^N}{\hat{y}^1,\cdots,\hat{y}^N}$$ where $y^i$ and $\hat{y}^i$ are the input and the output of the downlink channel from $\text{USN}_{\pi\paren{i}}$ to the $i$th player, respectively, $Y=\left[y^1,\cdots,y^N\right]^\top$, $p_{Y}\paren{\vec{y}}$ is the joint distribution of $Y$, and $\alpha$ is the total  average power constraint of the downlink channels between USNs and players.

 \begin{definition}\label{Def: 2epsilon}
A subset of $\mathcal{S}$ is $2\epsilon$-distinguishable if the distance between any two of its  points is more than $2\epsilon$ \cite{KT59}. 
\end{definition}
\begin{definition}
Let $\mathcal{M}_{2\epsilon}\paren{\mathcal{S}}$ denote the cardinality of maximal size $2\epsilon$ distinguishable subsets of $\mathcal{S}$. Then, the Kolmogorov capacity of $\mathcal{S}$ is defined as $\log\mathcal{M}_{2\epsilon}\paren{\mathcal{S}}$ \cite{KT59}.
\end{definition}

The next theorem establishes a lower bound on $T^\star_{\epsilon,\delta}\paren{\mathcal{G},\mathcal{O}}$. 
\begin{theorem}\label{Theo: DownCap}
Let $T^\star_{\epsilon,\delta}\paren{\mathcal{G},\mathcal{O}}$ denote the  complexity of the class of $N$-player non-cooperative games $\mathcal{G}$ with the continuous action space $\mathcal{S}$. Then, we have  
\begin{eqnarray}
T^\star_{\epsilon,\delta}\paren{\mathcal{G},\mathcal{O}} \geq\frac{\paren{1-\delta}\log\mathcal{M}_{2\epsilon}\paren{\mathcal{S}}-1}{C_{\rm down}}
\end{eqnarray}
where $C_{\rm down}$ is the total capacity downlink channels \textcolor{black}{from USNs to players}, and $\log\mathcal{M}_{2\epsilon}\paren{\mathcal{S}}$ is the Kolmogorov $2\epsilon$-capacity of the action space $\mathcal{S}$.% \cite{KT59} and $\mathcal{M}_{2\epsilon}\paren{\mathcal{S}}$ is the cardinality of maximal size, $2\epsilon$-distinguishable subsets of $\mathcal{S}$. 
\end{theorem}
\begin{IEEEproof}
See Subsection \ref{Proof: DownCap}.
\end{IEEEproof}
Theorem \ref{Theo: DownCap} establishes an algorithm-independent lower bound on the order of complexity of solving the game class $\mathcal{G}$. According to this theorem, $T^\star_{\epsilon,\delta}\paren{\mathcal{G},\mathcal{O}}$ is lower bounded by the ratio of the Kolmogorov $2\epsilon$-capacity of the action space $\mathcal{S}$ to the total Shannon capacity of the downlink channels. Note that the Kolmogorov $2\epsilon$-capacity of $\mathcal{S}$ can be interpreted as a measure of players' ambiguity about their NE strategies. Thus, as $\log\mathcal{M}_{2\epsilon}\paren{\mathcal{S}}$ becomes large, $T^\star_{\epsilon,\delta}\paren{\mathcal{G},\mathcal{O}}$ is expected to increase since players have to search in a bigger space to find their NE strategies. Based on Theorem \ref{Theo: DownCap}, $C_{\rm down}$ has a reverse impact on $T^\star_{\epsilon,\delta}\paren{\mathcal{G},\mathcal{O}}$. Note that $C_{\rm down}$ is an indication of the information transmission quality from USNs to players. That is, as  $C_{\rm down}$ decreases, players will receive noisier information regarding their utility functions compared with a large value of $C_{\rm down}$. %Thus, the complexity of solving game increases as players require to submit more queries to remove the impact of noisy communication channels on NS algorithm.
%Finally, we note that this result is independent of the computation model.
%{\bf TBD: $\log\mathcal{M}_{2\epsilon}\paren{\mathcal{S}}$ is called Kolmogorov $2\epsilon$ capacity of the action space \cite{KT59}. It is interpretation and the impact of $\epsilon$ on the complexity. We can also introduce $\log\mathcal{M}_{2\epsilon}\paren{\mathcal{S}}$ as the uncertainty capacity or a measure of uncertainty about the solution.}

\textcolor{black}{Theorem \ref{Theo: DownCap} depends on the $2\epsilon$-capacity of the constraint set $\mathcal{S}$ which is usually hard to compute unless the action space of players is restricted to special geometries. As $\mathcal{M}_{2\epsilon}\paren{\mathcal{S}}$ is just the maximum number of $\epsilon$-balls that can be packed into $\mathcal{S}$, it is asymptotically equal to $\frac{{\rm Vol}\paren{\mathcal{S}}}{{\rm Vol}\paren{B_\epsilon}} = \frac{{\rm Vol}\paren{\mathcal{S}}}{\alpha_N\epsilon^N}$ as $\epsilon$ tends to zero, where $B_\epsilon$ is the $N$-ball of radius $\epsilon$, and $\alpha_N$ is the $N$-dimensional spherical constant under the assumed norm. Thus, the complexity is at least of order $\log\frac{1}{\epsilon}$ as $\epsilon$ becomes small. The next result establishes a non-asymptotic lower bound of the same order, by lower bounding $\mathcal{M}_{2\epsilon}\paren{\mathcal{S}}$ using a result from lattice theory.} %This allows us to observe the impacts of $\epsilon$, volume of $\mathcal{S}$ and the surface area of $\mathcal{S}$ on the complexity of solving non-cooperative games. 
\begin{corollary}\label{Cor: DownCap}
The complexity of solving the class of $N$-player non-cooperative games $\mathcal{G}$ with continuous action space $\mathcal{S}$ can be bower bounded as 
\begin{eqnarray}
T^\star_{\epsilon,\delta}\paren{\mathcal{G},\mathcal{O}}\geq \frac{\paren{1-\delta}\!\!\paren{N\log\frac{1}{2\epsilon}\!+\!\logp{{\rm Vol}\paren{\mathcal{S}}\!-\!\epsilon{\rm P}\paren{\mathcal{S}}}}-1}{C_{\rm down}}
\end{eqnarray}
where ${\rm Vol}\paren{\mathcal{S}}$ and ${\rm P}\paren{\mathcal{S}}$ are the volume and the surface area of the action space of players, respectively. 
\end{corollary}
\begin{IEEEproof}
See Subsection \ref{Proof: Cor-DownCap}.
\end{IEEEproof}
Based on Corollary \ref{Cor: DownCap}, the lower bound on $T^\star_{\epsilon,\delta}\paren{\mathcal{G},\mathcal{O}}$ increases at least linearly with the number of players. This is due to the fact that the amount of uncertainty about the NE increases as the number of players becomes large. Recall that $\log\mathcal{M}_{2\epsilon}\paren{\mathcal{S}}$ is a quantitative indicator of ambiguity about the NE. Furthermore, $\epsilon$ has a logarithmic effect on $T^\star_{\epsilon,\delta}\paren{\mathcal{G},\mathcal{O}}$, \emph{i.e.,} the complexity of solving the class of games $\mathcal{G}$ increases according to $\Omega\paren{\log\frac{1}{\epsilon}}$ as $\epsilon$ becomes small. Hence, based on Corollary \ref{Cor: DownCap}, the game class $\mathcal{G}$ cannot be solved faster than $\Theta\paren{\log\frac{1}{\epsilon}}$ time-steps regardless of uplink/downlink noise distributions, and the computation model at the USNs.

According to Corollary \ref{Cor: DownCap}, the lower bound on the complexity of solving the game class $\mathcal{G}$ increases as the volume of the action space of players becomes large. Also, for a given surface area of action space of players, \emph{i.e.,} ${\rm P}\paren{\mathcal{S}}$, the volume of action space of players can be upper bounded using the isoperimetric inequality for convex bodies \cite{Gardner02} as follows:
\begin{eqnarray}\label{Eq: IICB}
{\rm Vol}\paren{\mathcal{S}}\leq \frac{{\rm Vol}\paren{\rm B}}{\paren{{\rm P}\paren{\rm B}}^{\frac{N}{N-1}}}{\rm P}\paren{\mathcal{S}}^{\frac{N}{N-1}}
\end{eqnarray}
where $\rm B$ is the closed unit ball in $N$-dimensional Euclidean space $\R^N$. Note that the equality in \eqref{Eq: IICB} is achieved if and only if $\mathcal{S}$ is a ball in $\R^N$\cite{Gardner02}. Thus, for a given surface area of action space of players ${\rm P}\paren{\mathcal{S}}$, the lower bound on the complexity of solving games in the class $\mathcal{G}$ increases as the action space of players becomes closer to a ball in $\R^N$ with the volume $\frac{{\rm Vol}\paren{\rm B}}{\paren{{\rm P}\paren{\rm B}}^{\frac{N}{N-1}}}{\rm P}\paren{\mathcal{S}}^{\frac{N}{N-1}}$.

%{\bf TBD: The impact of number of players on complexity}
%Based on Corollary \ref{Cor: DownCap}, the lower bound on $T^\star_{\epsilon,\delta}\paren{\mathcal{O}}$ increases at least linearly with the number of players. This is due to the fact that the amount of uncertainty about the NE increases as the number of players becomes large. Recall that $\log\mathcal{M}_{2\epsilon}\paren{\mathcal{S}}$ is a quantitative indicator of ambiguity about the NE, and it is a non-decreasing function of the number of players. Furthermore, $\epsilon$ has a logarithmic effect on $T^\star_{\epsilon,\delta}\paren{\mathcal{O}}$, \emph{i.e.,} the complexity of solving the class of games $\mathcal{G}$ increases according to $\Omega\paren{\log\frac{1}{\epsilon}}$ as $\epsilon$ becomes small. {\bf Regardless of noise distribution cannot be solved faster than exponentially}. %This is due to the fact that  the $\log\mathcal{M}_{2\epsilon}\paren{\mathcal{S}}$ is a non-decreasing function of $\epsilon$.

%{\bf TBD: The impact of $\epsilon$ on complexity}
%Theorem \ref{Theo: DownCap} and Corollary \ref{Cor: DownCap} only show the impacts of the total downlink capacity of players and Kolmogorov $2\epsilon$-capacity of action space of players on the complexity of solving non-cooperative games. 
\subsection{Partial-derivative Computation Model at USNs and Gaussian Uplink/Downlink Channels }\label{Subsec: Gauss}
In this section, we establish a lower bound on the complexity of solving a subclass of $\mathcal{G}$, denoted by $\mathcal{G}_\gamma$, under the partial-derivative computation model (see equation \eqref{Eq: Der-only})  when the communication noise in the uplink and downlink channels is Gaussian distributed. We also compare the complexity of solving the game class $\mathcal{G}_\gamma$ with the complexity of solving the class of strongly convex optimization problems. To specify the game class $\mathcal{G}_\gamma$, we first define the notion of pseudo-gradient for a utility function vector as follows.
\begin{definition} 
The pseudo-gradient of the utility function vector $U\paren{\vec{x}}=\left[u_1\paren{x^1,\vec{x}^{-1}},\cdots,u_N\paren{x^{N},\vec{x}^{-N}}\right]^\top$ is defined as 
\begin{eqnarray}
\tilde{\nabla}U\paren{\vec{x}}\!\!=\!\!\left[\frac{\partial }{\partial \paren{x^1}}u_1\paren{x^1,\vec{x}^{-1}},\cdots,\frac{\partial }{\partial \paren{x^N}}u_N\paren{x^N,\vec{x}^{-N}}\right]^\top\nonumber
\end{eqnarray}
 %We also restrict our attention to a subclass of $\mathcal{G}$. To define this subclass, let $\tilde{\nabla}U\paren{\vec{x}}=\left[\frac{\partial }{\partial \paren{x^i}}u_i\paren{x^i,\vec{x}^{-i}}\right]_i$ denote the pseudo-gradient induced by the utility functions of players. Also, let 
We use $J_{\tilde{\nabla}U}\paren{\vec{x}}$ to denote the Jacobian matrix of the vector valued function $\tilde{\nabla}U\paren{\vec{x}}$, \emph{i.e.,} 
\begin{eqnarray}
\left[J_{\tilde{\nabla}U}\paren{\vec{x}}\right]_{ij}=\frac{\partial^2 }{\partial \paren{x^j} \paren{x^i}}u_i\paren{x^i,\vec{x}^{-i}}, \quad 1\leq i,j\leq N\nonumber
\end{eqnarray}
\end{definition}
We  next specify a set of utility vector functions, denoted by $\mathcal{F}_\gamma$, which is used to define the game class $\mathcal{G}_\gamma$.
\begin{definition}
The set of utility vector functions $\mathcal{F}_\gamma$ is defined as the set of all vector valued functions $U\paren{\vec{x}}$ from $\R^N$ to $\R^N$ such that
\begin{enumerate}
\item The $N$-player non-cooperative game with utility vector function given by $U\paren{\vec{x}}$ and the constraint set $\mathcal{S}$ admits at least a Nash equilibrium (NE). 
\item The matrix  $J_{\tilde{\nabla}U}\paren{\vec{x}}$ exists for all $\vec{x}$ in $\mathcal{S}$.
\item The matrix  $J_{\tilde{\nabla}U}\paren{\vec{x}}$ satisfies $\norm{J_{\tilde{\nabla}U\paren{\vec{x}}}}{}\geq \gamma>0$ for all $\vec{x}\in\mathcal{S}$ where $\norm{J_{\tilde{\nabla}U}\paren{\vec{x}}}{}$ denotes  the matrix norm of $J_{\tilde{\nabla}_U}\paren{\vec{x}}$.
\end{enumerate}
\end{definition}
The next definition specifies the game class $\mathcal{G}_\gamma$.
\begin{definition} 
The class of games $\mathcal{G}_\gamma=\langle \mathcal{N}, \mathcal{S},\mathcal{F}_\gamma\rangle$ is defined as the set of all non-cooperative games with $N$ players, the constraint set $\mathcal{S}$, and the utility function vector $U\paren{\cdot}$ in $\mathcal{F}_\gamma$. 
\end{definition}

Note that the game class $\mathcal{G}_\gamma$ reduces  to the class $\mathcal{G}$ when $\gamma$ is equal to zero. The complexity of solving the game class $\mathcal{G}_\gamma$ heavily depends on $J_{\tilde{\nabla}U}\paren{\vec{x}}$ as shown in Theorem \ref{Theo:  Gaussian-bound}.

\begin{remark}
We note that both $\tilde{\nabla}U\paren{\vec{x}}$ and $J_{\tilde{\nabla}U}\paren{\vec{x}}$ play important roles in the game theory and system theory literature. To clarify this point, consider an unconstrained $N$-player game with the utility vector function $U\paren{\vec{x}}=\left[u_i\paren{x^i,\vec{x}^{-i}}\right]_i$ such that each $u_i\paren{x^i,\vec{x}^{-i}}$ is concave in $x^i$. Then, any solution of $\tilde{\nabla}_U\paren{\vec{x}}=\vec{0}$ will be a NE of this game. Also, consider the dynamical system $\dot{\vec{x}}=\tilde{\nabla}U\paren{\vec{x}}$. % where $\dot{\vec{x}}=\left[\frac{dx^i\paren{t}}{dt}\right]_i$.
 Then, any NE of the aforementioned game will be an equilibrium of this dynamical system and the eigenvalues of the matrix $J_{\tilde{\nabla}U}\paren{\vec{x}}$ determine the local stability of this dynamical system around its equilibria. Moreover, the matrix $J_{\tilde{\nabla}U}\paren{\vec{x}}$ can be used to study the uniqueness of the NE in non-cooperative games \cite{Rosen65}.% {\bf Let $\vec{x}_{\rm NE}$ be a locally stable equilibrium of this dynamical system. Then, the behavior of the dynamical system $\dot{\vec{x}}=\tilde{\nabla}U\paren{\vec{x}}$ around this equilibrium can be approximated by the behavior of its linearized $\dot{\vec{x}}=J_{\tilde{\nabla}U}\paren{\vec{x}_{\rm NE}}\vec{x}$ around $\vec{x}=0$. .}
\end{remark}

The next theorem studies the complexity of solving the game class $\mathcal{G}_\gamma$ under the partial-derivative computation model and Gaussian distributed uplink/downlink channels.  In the derivation of Theorem \ref{Theo:  Gaussian-bound}, it is assumed that the constraint set $\mathcal{S}$ contains a 2-ball with radius $\sqrt{2}\epsilon$, \emph{i.e.,} the set of all points  in a 2-dimensional plane with the distance $\sqrt{2}\epsilon$ from a point in $\mathcal{S}$.
\begin{theorem}\label{Theo:  Gaussian-bound}
Let $T^\star_{\epsilon,\delta}\paren{\mathcal{G}_\gamma,\mathcal{O}^1}$ denote the complexity of solving the game class $\mathcal{G}_{\gamma}$ under the partial-derivative computation model at USNs. Then, for Gaussian distributed uplink/downlink channels and $\delta\leq 0.5$, we have  
\begin{eqnarray}
T^\star_{\epsilon,\delta}\paren{\mathcal{G}_\gamma,\mathcal{O}^1}\geq  \frac{\paren{2\paren{1-\delta}-1}\min_i\sigma^2_i}{4\gamma^2\epsilon^2},
\end{eqnarray}
\textcolor{black}{where $\sigma_i^2$ is the variance of noise at the player $i$'s receiver.}
\end{theorem}
\begin{IEEEproof}
See Subsection \ref{Proof:  Gaussian-bound}.
\end{IEEEproof}
Theorem \ref{Theo:  Gaussian-bound} establishes an algorithm-independent lower bound on the complexity of solving the game class $\mathcal{G}_\gamma$. According to this result, the game class $\mathcal{G}_\gamma$ cannot be solved faster than $\Theta\paren{\frac{1}{\gamma^2\epsilon^2}}$ under the partial-derivative computation model and Gaussian noise model for uplink and downlink channels.

It is helpful to compare the complexity of solving the game class $\mathcal{G}_\gamma$ with that of solving a black-box convex optimization problem. The complexity of solving black-box optimization problems is studied under an oracle-based setting in which  optimization algorithms  rely on an oracle for function evaluation. In this setting, the oracle receives noise-free queries from the optimization algorithm, and, the algorithm receives a noisy version of oracle's response \cite{RR11}. The next corollary studies the complexity of solving the game class $\mathcal{G}_\gamma$ in a similar setting, \emph{i.e.,} under noiseless uplink channels and Gaussian downlink channels.
\begin{corollary}\label{Cor: Gaussian}
Let $T^\star_{\epsilon,\delta}\paren{\mathcal{G}_\gamma,\mathcal{O}^1}$ denote the complexity of solving the class $\mathcal{G}_{\gamma}$ using the partial-derivative computation model. Then, for noiseless uplink channels, Gaussian distributed downlink channels and $\delta\leq 0.5$, we have  
\begin{eqnarray}
T^\star_{\epsilon,\delta}\paren{\mathcal{G}_\gamma,\mathcal{O}^1}\geq  \frac{\paren{2\paren{1-\delta}-1}\min_i\sigma^2_i}{4\gamma^2\epsilon^2}.\nonumber
\end{eqnarray}
\end{corollary} 
\begin{IEEEproof}
The proof is similar to that of Theorem \ref{Theo:  Gaussian-bound} and is skipped.
\end{IEEEproof}
Next, we use Corollary \ref{Cor: Gaussian} to compare the complexity of solving the class $\mathcal{G}_\gamma$ with the complexity of solving a class of convex optimization problems using the oracle-based setting. To this end, consider the following optimization problem 
  \begin{eqnarray} 
\underset{\vec{x}\in\mathcal{S}}{\min} & f\paren{\vec{x}},\nonumber 
 \end{eqnarray}
where $\mathcal{S}$ is a convex set, and $f\paren{\vec{x}}$ belongs to the class of continuous and strongly convex functions $\mathcal{F}_{\rm sc}$. The complexity of solving the class of convex optimization problems with the objective function in $\mathcal{F}_{\rm sc}$ is defined as \cite{RR11}
\begin{align}
&\inf\left\{T^{}\in\N: \exists \mathcal{A}\quad s.t. \sup_{f\paren{\cdot}\in \mathcal{F}_{\rm sc}}\PRP{f\paren{\vec{x}_{T+1}}-f^\star\geq \epsilon}\leq \delta\right\}\nonumber
\end{align}
where $\vec{x}_{T+1}$ is the output of the algorithm $\mathcal{A}$ after $T$ queries, and $f^\star=\inf_{\vec{x}\in\mathcal{S}}f\paren{\vec{x}}$. It is shown in \cite{RR11} that the complexity of solving the class of strictly convex optimization problems under the subgradient computation model and Gaussian noise model is given by $\Omega\paren{\frac{1}{\epsilon}}$. According to the Corollary \ref{Cor: Gaussian}, the game class $\mathcal{G}_{\gamma}$ is harder to solve compared with the class of strictly convex optimization problems since the games are non-convex problems, and NE is more sophisticated solution concept compared with  the minimizer of a convex function (see Remark \ref{Remark: NE} for more details).
\subsection{Partial-derivative Computation Model at USNs, Non-Gaussian Downlink Channels and Noiseless Uplink Channels}\label{Subsec: Non-gaussian}
In this subsection, we study the complexity of solving the game class $\mathcal{G}_\gamma$ under the partial-derivative computation model when the downlink channels are not \emph{necessarily} Gaussian and the uplink channels are noiseless.  To this end, let  $p_{V^i}\paren{x}$ denote the \emph{common} probability distribution function (PDF) of the collection of random variables $\left\{V^i_k\right\}_k$, \emph{i.e.,} the collection of noise terms in the downlink channel from ${\rm USN}_{\pi\paren{i}}$ to player $i$. %Also, let $V^i$ denote a generic random variable distributed according to $p_{V^i}\paren{x}$. Then, for \emph{each} player $i$, we assume that 
To investigate the complexity of the game class $\mathcal{G}_\gamma$ in the non-Gaussian setting, we assume that $p_{V^i}\paren{x}$ satisfies the following mild assumptions for all $1\leq i\leq N$
\begin{enumerate}
\item The PDF $p_{V^i}\paren{x}$ is non-zero everywhere on $\R$.
\item The PDF $p_{V^i}\paren{x}$ is at least 3 times continuously differentiable, \emph{i.e.,} $p_{V^i}\paren{x}\in C^3$.
\item There exist positive constants $\beta_1,\beta_2,\beta_3>0$ such that 
\begin{eqnarray}
\abs{\frac{d^3}{dx^3}\log p_{V^i}\paren{x}}\leq \beta_1+\beta_2 \abs{x}^{\beta_3}\quad \forall x \in \R\nonumber
\end{eqnarray}
\item The tail of the random variable $\abs{V^i_k}$ decays faster than $x^{-\paren{\beta_3+1}}$, \emph{i.e.,} we have 
\begin{eqnarray}
\lim_{x\rightarrow\infty}x^{\paren{\beta_3+1+r}}\PR{\abs{V^i_k}\geq x }=0\quad \forall k\nonumber
\end{eqnarray}
 for some $r>0$.
\end{enumerate}  
The next theorem derives a lower bound on the complexity of solving the game class $\mathcal{G}_\gamma$ in the non-Gaussian setting.
\begin{theorem}\label{Theo:  Non-Gaussian-bound}
Let $T^\star_{\epsilon,\delta}\paren{\mathcal{G}_\gamma,\mathcal{O}^1}$ denote the complexity of solving the $N$-player non-cooperative games in the class $\mathcal{G}_{\gamma}$ using the partial-derivative computation model at USNs. Assume that the PDFs of the downlink noise terms satisfy the assumptions 1-4 above and the uplink channels are noiseless.  Then, for $\delta\leq 0.5$ we have  
\begin{eqnarray}
T^\star_{\epsilon,\delta}\paren{\mathcal{G}_\gamma,\mathcal{O}^1}\geq\frac{2\paren{1-\delta}-1}{4N\epsilon^2\gamma^2\max_i\mathcal{I}_i+\BO{\epsilon^3}}.\nonumber
\end{eqnarray}
\textcolor{black}{where $\mathcal{I}_i$ is the Fisher information of the PDF $p_{V^i}\paren{x}$  with respect to a shift parameter.}
\end{theorem}
\begin{IEEEproof}
See Subsection \ref{Proof: Gen-noise}.
\end{IEEEproof}
Theorem \ref{Theo:  Non-Gaussian-bound} establishes a lower bound on the complexity of solving the game class $\mathcal{G}_\gamma$ under noiseless uplink channels and non-Gaussian downlink channels. In  Corollary \ref{Cor: Gaussian}, we showed that the complexity of solving the game class $\mathcal{G}_\gamma$ under the partial-derivative computation model is at least of the order ${\frac{1}{\gamma^2\epsilon^2}}$, \textcolor{black}{as $\epsilon$ becomes small}, when the uplink channels are noiseless and the downlink channels are Gaussian distributed. According to Theorem \ref{Theo:  Non-Gaussian-bound}, \textcolor{black}{the lower bound on the complexity of solving the game class $\mathcal{G}_\gamma$ is also of the order $\frac{1}{\gamma^2\epsilon^2}$ when  the uplink channels are noiseless and the downlink channels are not necessarily Gaussian distributed.} 

Theorem \ref{Theo:  Non-Gaussian-bound} is established by deriving an asymptotic expansion for the Kullback-Leibler (KL) distance between the PDF $p_{V^i}\paren{x}$ and its shifted version. More precisely, we show that 
\begin{align}
&\KLD{p_{V^i}\paren{x}}{p_{A_i\vec{\tau}+V^i}\paren{x}}= \frac{1}{2}\mathcal{I}_i\paren{A_i\vec{\tau}}^2+\BO{\norm{\vec{\tau}}{}^3}\nonumber
\end{align}
where $A_i$ is a 1-by-$N$ row vector, $\vec{\tau}$ is an $N$-by-1 column vector, $p_{A_i\vec{\tau}+V^i}\paren{x}$ is the PDF of $A_i\vec{\tau}+V^i$, and $\mathcal{I}_i$ is the Fisher information of $p_{V^i}\paren{x}$ with respect to a shift parameter. \textcolor{black}{Since the Taylor series of a real function is not necessarily convergent, Theorem \ref{Theo:  Non-Gaussian-bound} is proved using Taylor expansion Theorem. The assumptions 1-4 above are used to bound the remainder integral which appears in the Taylor expansion (see Lemma \ref{Lem: KL-expansion} in Subsection \ref{Proof: Gen-noise} and its proof for more details).}
\subsection{Partial-derivative Computation Model at USNs With Arbitrarily Distributed Uplink and Downlink Channels}\label{Sec: Gen-up-down}
The next theorem establishes a lower bound  on the complexity of solving the game class $\mathcal{G}_\gamma$ under the partial-derivative computation model when the uplink and downlink channels are not necessarily Gaussian distributed. %In this section, we assume that the Lebesgue measure on  probability measure of each uplink/downlink noise term and satisfies the (related) assumptions in Subsection \ref{Subsec: Assum}.
\begin{theorem}\label{Theo: MI-bound}
The complexity of solving the game class $\mathcal{G}_\gamma$ using the partial-derivative computation model at USNs is lower bounded by 
\begin{eqnarray}\label{Eq: MI-Theo}
T^\star_{\epsilon,\delta}\paren{\mathcal{G}_\gamma,\mathcal{O}^1}\geq\sup_{A:\norm{A}{}\geq \gamma,{\mathcal{S}}_{2\epsilon}\in\mathcal{S}} \frac{\paren{1-\delta}\log\abs{{\mathcal{S}}_{2\epsilon}}-1}{\MI{\vec{x}^\star_M}{-A\vec{x}^\star_M+\hat{W}_A}}
\end{eqnarray}
where ${\mathcal{S}}_{2\epsilon}$ is a $2\epsilon$-distinguishable subset of $\mathcal{S}$, $A=\left[a_{ij}\right]$ is an $N$-by-$N$ symmetric, negative definite matrix, $\vec{x}^\star_M$ is a random vector taking value in ${\mathcal{S}}_{2\epsilon}$ with uniform distribution, the random vector $\hat{W}_A=\left[\hat{W}^i_1\right]_{i=1}^N$ is defined as   
\begin{align}
\hat{W}^i_1=\paren{\sum_{j\in\mathcal{N}_{{\rm usn}_{\pi\paren{i}}}}a_{ij}W^{j}_{1,{\rm usn}_{\pi\paren{i}}}}+V^i_1,\quad 1\leq i\leq N\nonumber
\end{align} 
and $\MI{\vec{x}^\star_M}{-A\vec{x}^\star_M+\hat{W}_A}$ is the mutual information between $\vec{x}^\star_M$ and $-A\vec{x}^\star_M+\hat{W}_A$. 
\end{theorem}
\begin{IEEEproof}
See Subsection \ref{Theo: MI-bound}.
\end{IEEEproof}
Theorem \ref{Eq: MI-Theo} derives a lower bound on the complexity of solving the game class $\mathcal{G}_\gamma$ which depends on the constraint set $\mathcal{S}$, the constant $\gamma$ and the noise distribution in the uplink and downlink channels. The optimization in \eqref{Eq: MI-Theo} is over the set of all symmetric, negative definite matrices with norm greater than or equal to $\gamma$, and the set of all $2\epsilon$-distinguishable subsets of $\mathcal{S}$. The matrix $A$ in \eqref{Eq: MI-Theo}  stems from the construction of quadratic utility functions in the proof of Theorem \ref{Theo: MI-bound}, the set $\mathcal{S}_{2\epsilon}$ and the matrix $A$ jointly represent a finite subset of the function class $\mathcal{F}_\gamma$, and $\hat{W}^i_1$ represents the combined impact of  uplink and downlink channels at player $i$'s receiver under the constructed quadratic utility functions (see the proof of this theorem for more details).  %In this theorem, we do not impose any distributions on the communication noises. 

 Theorem \ref{Theo: MI-bound} can be used to numerically obtain a lower bound on the complexity of solving the game class $\mathcal{G}_\gamma$ up to $\epsilon$ accuracy when the uplink/downlink channels are not Gaussian distributed. Note that according to \eqref{Eq: MI-Theo}, $T^\star_{\epsilon,\delta}\paren{\mathcal{G}_\gamma,\mathcal{O}^1}$ can be lower bounded as 
\begin{eqnarray}\label{Eq: MI-Num}
T^\star_{\epsilon,\delta}\paren{\mathcal{G}_\gamma,\mathcal{O}^1}\geq \frac{\paren{1-\delta}\log\abs{{\mathcal{S}}_{2\epsilon}}-1}{\MI{\vec{x}^\star_M}{-A\vec{x}^\star_M+\hat{W}_A}}
\end{eqnarray}
% Although, the lower bound in Theorem \ref{Theo: MI-bound} is {\bf difficult} to evaluate, it allows to derive a lower bound on the complexity of solving a class of games when the communication noises are Gaussian distributed in the next subsection.  
where $A$ is a symmetric, negative definite matrix with $\norm{A}{}\geq \gamma$ and $\mathcal{S}_{2\epsilon}$ is a $2\epsilon$-distinguishable subset of $\mathcal{S}$. Thus, by numerically evaluating the mutual information term in \eqref{Eq: MI-Num}, one can obtain a lower bound on $T^\star_{\epsilon,\delta}\paren{\mathcal{G}_\gamma,\mathcal{O}^1}$.
%\begin{remark}
%{\bf The derivation of lower bound in Theorem \ref{Theo:  Gaussian-bound} relies on the expression for the Kullback-Leibler (KL) distance between two Gaussian probability density functions (PDFs). However, this technique may not be applicable when the uplink and downlink noise terms are not Gaussian distributed as one may not find a closed from expression for the KL distance between two non-Gaussian PDFs.}
%\end{remark}

The lower bound in Theorem \ref{Theo: MI-bound} has the following information-theoretic interpretation.  Consider an auxiliary multiple-input-single-output (MISO) broadcast channel with $\vec{x}^\star_M$ as input and  the $-A\vec{x}^\star_M+\hat{W}_A$ as output. Here, the channel input, \emph{i.e.,} $\vec{x}^\star_M$, takes value from the finite set of input alphabets ${\mathcal{S}}_{2\epsilon}$ with uniform distribution. The symmetric, positive definite matrix $-A$ acts on the input, and the received signal by player $i$ is given by $-A_i\vec{x}^\star_M+\hat{W}_1^i$ where $A_i$ is the $i$th row of $A$. Note that $\log\abs{{\mathcal{S}}_{2\epsilon}}$ can be intuitively interpreted as the transmitter's bit-rate and $\MI{\vec{x}^\star_M}{-A\vec{x}^\star_M+\hat{W}_A}$ can be  intuitively deemed as the amount of common information between the transmitted signal and the set of received signals by players. Therefore, 
\begin{align}
{\rm R}\paren{A,\mathcal{S}_{2\epsilon}}=\frac{\MI{\vec{x}^\star_M}{-A\vec{x}^\star_M+\hat{W}_A}}{\paren{1-\delta}\log\abs{{\mathcal{S}}_{2\epsilon}}-1}\nonumber
\end{align}
can be viewed as the relative common information between the transmitted signal and the set of received signals by players for a particular choice of the set ${\mathcal{S}}_{2\epsilon}$ and the matrix $A$. Note that $\MI{\vec{x}^\star_M}{-A\vec{x}^\star_M+\hat{W}_A}\leq \DSE{\vec{x}^\star_M}=\log\abs{{\mathcal{S}}_{2\epsilon}}$ as $\vec{x}^\star_M$ is uniformly distributed over ${\mathcal{S}}_{2\epsilon}$.  Thus, according to \eqref{Eq: MI-Theo}, the complexity of solving the game class $\mathcal{G}_\gamma$ is limited by %the least efficient axillary MISO channel induced by the set ${\mathcal{S}}_{2\epsilon}$ and $A$, \emph{i.e.,} 
the choice of ${\mathcal{S}}_{2\epsilon}$ and $A$ such that the transmitted signal and the  set of received signals by players have the smallest amount of relative common information. 

%{\bf Noise will not depend on $A$ when the uplink noise is zero}
%\begin{enumerate}
% \item $\log\abs{\mathcal{S}_{2\epsilon}}$ is the nominal data-rate.
%\item $\MI{\vec{x}_M}{-A\vec{x}_M+\hat{W}_A}$ actual data-rate. 
%\item $A$ is the Jacobin of the "gradient flow" induced by the utilities of players around NE.
%\item The right hand side of \eqref{Eq: MI-Theo} is the inefficiency of the MIMO communication system induced by the Jacobian of gradient flow and $\mathcal{S}_{2\epsilon}$, \emph{i.e.,} the $2\epsilon$-distinguishable subset of the $\mathcal{S}$.
%\item The matrix $A$ controls the channel gains and the noise at the receiver side of the MIMO system, whereas $\mathcal{S}_{2\epsilon}$ controls the input alphabet.
%\end{enumerate}
%%%%%%%%%%%%%%%%%%%%%%%%%%%%%%%%%%%%%%%%%%%%%%%%%%%%%%%%%%%%%%%%%%%%%%%%%%%%%%%%%%%%%%%%%%%%%%%%%%%%%%%%
\section{Derivations of Results}\label{Sec: Derivations}
\subsection{Proof of Theorem \ref{Theo: DownCap}}\label{Proof: DownCap}
The proof of Theorem \ref{Cor: DownCap} is based on the following four steps:
\begin{enumerate}
 \item Firstly, we construct a finite subset of $\mathcal{F}$, denoted by $\mathcal{F}^\prime$%, such that the NE corresponding to two distinct elements of $\mathcal{F}^\prime$  are at least $2\epsilon$ apart 
(see subsection \ref{Subsec: Restrict}  for more details).
\item Secondly, for the function class $\mathcal{F}^\prime$, the Nash seeking problem is reduced to a hypothesis test problem %we lower bound the deviation probability of any algorithm $\mathcal{A}$ from the NE by the error probability of a hypothesis test % which operates based on the algorithm $\mathcal{A}$ 
 (see subsection \ref{Subsec: HT} for more details). 
\item Thirdly, the generalized Fano inequality is used to obtain a lower bound on the error probability of the hypothesis test problem (see subsection \ref{Subsec: HT} for more details). 
\item  Finally, information-theoretic inequalities are used to obtain an upper bound on the mutual information term which appears in the generalized Fano inequality. %such  that final lower bound on $T^\star_{\epsilon,\delta}\paren{\mathcal{O}}$ is independent of the algorithm $\mathcal{A}$.
\end{enumerate}

 \subsubsection{Restricting the Class of Utility Function Vectors}\label{Subsec: Restrict}
The first step in deriving the lower bound on $T^\star_{\epsilon,\delta}\paren{\mathcal{G},\mathcal{O}}$ is to restrict  our analysis to an appropriately chosen, finite subset of $\mathcal{F}$. %The set $\mathcal{F}^\prime$ can be represented as an indexed family of utility functions in $\mathcal{F}$, \emph{i.e.,} $\mathcal{F}^\prime=\left\{U_m\paren{\cdot}\in\mathcal{F}, m=1,\cdots,\abs{\mathcal{F}^\prime}\right\}$.  %The collection of elements of $\mathcal{F}^\prime$ are denoted by $\left\{U_i\paren{\cdot}\right\}_{i=1}^{\abs{\mathcal{F}^\prime}}$. 
%The set $\mathcal{F}^\prime$ is constructed as follows. %is given such that the Nash equilibria corresponding to different elements of $\mathcal{F}^\prime$ are at least $2\epsilon$ apart. 
To this end, let  
\begin{eqnarray}
\mathcal{S}^\star_{2\epsilon}=\left\{\vec{x}^\star_m\in\mathcal{S}: m=1,\cdots,\mathcal{M}_{2\epsilon}\paren{\mathcal{S}}\right\}
\end{eqnarray}
 be a maximal size, $2\epsilon$-distinguishable subset of $\mathcal{S}$ where $\mathcal{M}_{2\epsilon}\paren{\mathcal{S}}$ is the cardinality of  maximal size, $2\epsilon$-distinguishable subsets of $\mathcal{S}$ (see Definition \ref{Def: 2epsilon} for more details on $2\epsilon$-distinguishable subsets of $\mathcal{S}$). Next, for each $\vec{x}^\star_m\in\mathcal{S}^\star_{2\epsilon}$ ($m=1,\cdots,\mathcal{M}_{2\epsilon}\paren{\mathcal{S}}$), we construct a utility function vector $U_m\paren{\vec{x}}$ such that $\vec{x}^\star_m$ is the NE of the non-cooperative game with $N$ players, utility function vector $U_m\paren{\vec{x}}$ and the action space $\mathcal{S}$.

The utility function vector $U_m\paren{\vec{x}}$ ($m=1,\cdots,\mathcal{M}_{2\epsilon}\paren{\mathcal{S}}$) is constructed as follows. Let $A=\left[a_{ij}\right]_{i,j}$ be a symmetric, negative definite $N$-by-$N$ matrix. Also, let $u_{m,i}\paren{x^i,\vec{x}^{-i}}=\frac{a_{ii}}{2}\paren{x^i}^2+x^i\paren{-A_i\vec{x}^\star_m+\sum_{j\neq i} a_{ij}x^j}$ denote the utility function of player $i$ where $A_i$ is the $i$th row of $A$. The utility function vector $U_m\paren{\vec{x}}$ is constructed as $U_m\paren{\vec{x}}=\left[u_{m,i}\paren{x^i,\vec{x}^{-i}}\right]_i^\top$. Let $\mathcal{F}^\prime$ be the finite set of utility function vectors defined as 
\begin{eqnarray}
\mathcal{F}^\prime=\left\{U_m\paren{\cdot}\in\mathcal{F}, m=1,\cdots,\mathcal{M}_{2\epsilon}\paren{\mathcal{S}}\right\}
\end{eqnarray}
Clearly, we have $\abs{\mathcal{F}^\prime}=\mathcal{M}_{2\epsilon}\paren{\mathcal{S}}$.

The next lemma shows that the utility function vector $U_m\paren{\vec{x}}$ belongs to the function class $\mathcal{F}$, \emph{i.e.,} the class of vector-valued functions from $\R^N$ to $\R^N$ such that any $N$-player non-cooperative game with the constraint set $\mathcal{S}$ and utility function vector in $\mathcal{F}$ admits at least one NE. 
	\begin{lemma}\label{Lem: Uniq}
	Consider the $N$-player non-cooperative game in which: $(i)$ the utility function of the $i$th player is given by $u_{m,i}\paren{x^i,\vec{x}^{-i}}=\frac{a_{ii}}{2}\paren{x^i}^2+x^i\paren{-A_i\vec{x}^\star_m+\sum_{j\neq i} a_{ij}x^j}$, $(ii)$ the action space of players is given by $\mathcal{S}$. Then, $\vec{x}^\star_m$ is a NE of the game among players, and we have $U_m\paren{\vec{x}}\in\mathcal{F}$.
	\end{lemma}
	\begin{IEEEproof}
To prove this result, we first show that $\vec{x}_m^\star$ is the NE of the unconstrained, $N$-player non-cooperative game with the utility function vector $U_m\paren{\vec{x}}$ as follows. Consider the non-cooperative game in which the utility function of player $i$ is given by $u_{m,i}\paren{x^i,\vec{x}^{-i}}$. Then, the best response of the $i$th player to $\vec{x}^{-i}$ is obtained by solving the following optimization problem:
 \begin{eqnarray} \label{Eq: Opt}
\begin{array}{cc}
\underset{x^i}{\max} & u_{m,i}\paren{x^i,\vec{x}^{-i}}
\end{array}
 \end{eqnarray}
where $u_{m,i}\paren{x^i,\vec{x}^{-i}}=\frac{a_{ii}}{2}\paren{x^i}^2+x^i\paren{-A_i\vec{x}^\star_m+\sum_{j\neq i} a_{ij}x^j}$. Note that $a_{ii}<0$ for $1\leq i\leq N$ as the matrix $A$ is negative definite. Thus, the objective function in \eqref{Eq: Opt} is strongly concave in $x^i$ and the optimization problem \eqref{Eq: Opt} admits a unique solution. The  best response of player $i$ to $\vec{x}^{-i}$ can be obtained using the first order necessary and sufficient optimality condition:
 \begin{eqnarray}
-A_i\vec{x}^\star_m+\sum\limits_{j=1}^{N}a_{ij}x^j=0\nonumber
 \end{eqnarray}
Note that any intersection of the best responses of players is a NE. Thus, the NE of the unconstrained game can be found by solving the following system of linear equations 
 \begin{eqnarray}\label{Eq: BR}
-A_i\vec{x}^\star_m\!\!+\!\!\sum\limits_{j=1}^{N}a_{ij}x^j=0, 1\leq i\leq N
 \end{eqnarray}
It can be easily verified that $\vec{x}^\star_m$ is a solution of \eqref{Eq: BR} which implies $\vec{x}^\star_m$ is a NE of the $N$-player, unconstrained non-cooperative game with the utility function vector $U_m\paren{\vec{x}}$. Since $\vec{x}^\star_m$ belongs to $\mathcal{S}$, it is also a NE of the $N$-player, non-cooperative game with the utility function vector $U_m\paren{\vec{x}}$ and the action space $\mathcal{S}$. Thus, $U_m\paren{\cdot}$ belongs to the function class $\mathcal{F}$.
	\end{IEEEproof}

Lemma \ref{Lem: Uniq} implies that $\mathcal{F}^\prime$ is a subset of $\mathcal{F}$. We refer to the class of $N$-player non-cooperative games with the utility function vectors in $\mathcal{F}^\prime$  and the action space $\mathcal{S}$ as $\mathcal{G}^\prime=\langle \mathcal{N}, \mathcal{S},\mathcal{F}^\prime\rangle$. Here, we make the technical assumption that each game in the game class $\mathcal{G}^\prime$ admits a unique NE. % $\vec{x}^\star_m$ is the unique Nash equilibrium of the game among players when the utility functions of players are given by $U_m\paren{\vec{x}}$. 
This assumption can be satisfied by imposing more structure on the constraint set $\mathcal{S}$, \emph{e.g.,} see \cite{Rosen65}. In this paper, we do not explicitly impose a specific requirement on the action space $\mathcal{S}$ to guarantee the uniqueness of NE for the games in $\mathcal{G}^\prime$ since these restrictions are only sufficient conditions (not necessary and sufficient) to guarantee the existence  of a unique NE.  %Then, $T^\star_{\epsilon,\delta}\paren{\mathcal{O}}$ is lower bounded by \eqref{Eq: LB}.

Now, for a given $\epsilon$ and $\delta$,  consider any algorithm $\mathcal{A}$ for which after $T$ time-steps, we have 
\begin{align}
\!\!\!\!\sup_{U\paren{\cdot}\in \mathcal{F}}\inf_i\PRP{\norm{\vec{x}_{{\rm NE}_i, U\paren{\cdot}}\!\!-\!\!\mathcal{A}_{T+1}\!\!\paren{{X}_{1:T},{\hat{Y}_{1:T}},\hat{Z}_{1:T}}}{}\!\geq \!\epsilon}\leq \delta.\nonumber
\end{align}
 Since $\mathcal{F}^\prime$ is a subset of $\mathcal{F}$ and any game in $\mathcal{G}^\prime$ admits a unique NE, we have 
\begin{align}\label{Eq: Subset-Bound}
\!\!\sup_{m=1,\cdots,\mathcal{M}_{2\epsilon}\paren{\mathcal{S}}}\!\!\!\!\!\!\PRP{\norm{\vec{x}^\star_{m}-\mathcal{A}_{T+1}\paren{{X}_{1:T},{\hat{Y}_{1:T}},\hat{Z}_{1:T}}}{}\geq \epsilon}\leq \delta.
\end{align}

\subsubsection{A Genie-aided Hypothesis Test}\label{Subsec: HT}
In this subsection, we construct a genie-aided hypothesis test as follows which operates based on the output of the algorithm $\mathcal{A}$. %Then, we show that, for  utility function vectors in $\mathcal{F}^\prime$, the probability of $\epsilon$ deviation of any algorithm $\mathcal{A}$ (from the NE) is lower bounded by the error probability of its corresponding hypothesis test. 
Consider a genie-aided hypothesis test in which, first, a genie selects a game instance from $\mathcal{G}^\prime$ uniformly at random. Let $\vec{x}^\star_M\in\mathcal{S}^\star_{2\epsilon}$ and  $U_M\paren{\cdot}\in\mathcal{F}^\prime$ denote the NE and the  utility function vector associated with the randomly selected game instance, respectively, where $M$ is a random variable uniformly distributed over the set $\left\{1,\cdots,{\mathcal{M}}_{2\epsilon}\paren{\mathcal{S}}\right\}$.% The NE corresponding to $U_M\paren{\cdot}$ is denoted by $\vec{x}^\star_M$. Next, USNs are notified about the $U_M\paren{\cdot}$.  

At time $k\in\left\{1,\cdots,T\right\}$,  the $i$th player updates its action using the algorithm $\mathcal{A}$ according to $x^i_k=\mathcal{A}^i_k\paren{X^i_{1:k-1},\hat{Y}^i_{1:k-1},\hat{Z}^i_{1:k-1}}$. % Also, at each time, PN responds to players based on the utility functions given by $U_M\paren{\cdot}$. 
 At time $T+1$, the genie %has access to  $\left\{X^i_{1:T},\hat{Y}^i_{1:T},\hat{Z}^i_{1:T}\right\}_{i=1}^N$ and 
estimates the NE according to the following decision rule:
\begin{eqnarray}\label{Eq: HT}
 \vec{\hat{x}}^\star=\arg\min_{x\in \mathcal{S}_{2\epsilon}^\star}\!\!\norm{x-\mathcal{A}_{T+1}\paren{X_{1:T},\hat{Y}_{1:T},\hat{Z}_{1:T}}}{}
\end{eqnarray}
where $\vec{\hat{x}}^\star\in\mathcal{S}^\star_{2\epsilon}$ is the closest elements of $\mathcal{S}^\star_{2\epsilon}$ to the output of algorithm. An error is declared if the error event 
\begin{eqnarray}
E_{\mathcal{A}}=\left\{\vec{x}^\star_{M}\neq\vec{\hat{x}}^\star\right\}\nonumber
\end{eqnarray}
 happens, that is, if the estimated NE is not equal to the true NE.  % Finally, the genie declares an error if the event $E_{\mathcal{A}}$ occurs.
%\begin{remark}
%Please note that the proposed hypothesis scheme only provide a mathematical tool for lower bounding the maximum deviation probability of an algorithm from the NE, and in practice, the utility functions are not randomly selected. 
%\end{remark}
The next lemma establishes an upper bound on the probability of the error event $E_{\mathcal{A}}$.
\begin{lemma}\label{Lem: HT}
Let $\PRP{E_{\mathcal{A}}}$ denote the error probability under the proposed genie-aided hypothesis test. Then, 
\begin{align}
\PRP{E_{\mathcal{A}}}\leq\sup_{m\in\left\{1,\cdots,\mathcal{M}_{2\epsilon}\paren{\mathcal{S}}\right\}}\!\!\!\!\!\!\!\PRP{\norm{\vec{x}^\star_{m}-\mathcal{A}_{T+1}\paren{{X}_{1:T},{\hat{Y}_{1:T}},\hat{Z}_{1:T}}}{}\geq \epsilon}\nonumber
\end{align}
where $\vec{x}^\star_{m}$ is the NE corresponding to the utility function vector $U_m\paren{\cdot}$.
\end{lemma}
\begin{IEEEproof}
We show that the error event $E_{\mathcal{A}}$ implies 
\begin{align}
\norm{\vec{x}^\star_{M}-\mathcal{A}_{T+1}\paren{X_{1:T},\hat{Y}_{1:T},\hat{Z}_{1:T}}}{}>\epsilon\nonumber
\end{align}
 by contraposition. That is, we show if the following inequality holds 
\begin{align}\label{eq: Assump}
\norm{\vec{x}^\star_{M}-\mathcal{A}_{T+1}\paren{X_{1:T},\hat{Y}_{1:T},\hat{Z}_{1:T}}}{}\leq \epsilon,
\end{align}
 then, we have  $\vec{\hat{x}}^\star=\vec{x}^\star_{M}$. %Note that $\norm{\vec{x}_{{\rm NE},M}-\mathcal{A}_{T+1}\paren{X_{1:T},Y_{1:T}}}{}\leq \epsilon$ implies $\norm{\vec{x}^i_{{\rm NE},M}-\mathcal{A}^i_{T+1}\paren{X^i_{1:T},Y^i_{1:T}}}{}\leq \epsilon$ for $1\leq i\leq N$. 
Assume that the inequality \eqref{eq: Assump} holds. For $\vec{x}^\star_{m}\neq \vec{x}^\star_{M}$, we have 
\begin{align}
2\epsilon&\stackrel{(a)}{<}\norm{\vec{x}^\star_{m}-\vec{x}^\star_{M}}{}\nonumber\\
&\leq \norm{\vec{x}^\star_{m}-\mathcal{A}_{T+1}\paren{X_{1:T},\hat{Y}_{1:T},\hat{Z}_{1:T}}}{}+\norm{\vec{x}^\star_{M}-\mathcal{A}_{T+1}\paren{X_{1:T},\hat{Y}_{1:T},\hat{Z}_{1:T}}}{}\nonumber\\
&<\norm{\vec{x}^\star_{m}-\mathcal{A}_{T+1}\paren{X_{1:T},\hat{Y}_{1:T},\hat{Z}_{1:T}}}{}+\epsilon\nonumber
\end{align}
where $(a)$ follows from the fact that $\vec{x}_m^\star$ and $\vec{x}_M^\star$ belong to the $2\epsilon$-distinguishable set $\mathcal{S}^\star_{2\epsilon}$. %of the set $\mathcal{F}^\prime$. %Thus, $\vec{x}_{{\rm NE},m}\neq \vec{x}_{{\rm NE},M}$, we have $\norm{\vec{x}^i_{{\rm NE},m}-\mathcal{A}^i_{T+1}\paren{X_{1:T},Y_{1:T}}}{}>\epsilon$ for at least one player. Let $i$ be the index of such player. 
Thus, $\vec{x}^\star_{m}$ cannot be the solution of the optimization problem \eqref{Eq: HT}.
Therefore, we have 
 \begin{align}
%\PRP{\left.E_{\mathcal{A}}\right|M}\leq\PRP{\norm{\vec{x}_{{\rm NE}, M}-\mathcal{A}_{T+1}\paren{{X}_{1:T},\hat{Y}_{1:T},\hat{Z}_{1:T}}}{}\geq \epsilon}\nonumber
\PRP{E_{\mathcal{A}}}&\leq\PRP{\norm{\vec{x}^\star_{M}-\mathcal{A}_{T+1}\paren{{X}_{1:T},\hat{Y}_{1:T},\hat{Z}_{1:T}}}{}\geq \epsilon}\nonumber\\
&=\ESI{M}{\CP{\norm{\vec{x}^\star_{M}-\mathcal{A}_{T+1}\paren{{X}_{1:T},{\hat{Y}_{1:T}},\hat{Z}_{1:T}}}{}\geq \epsilon}{M}}\nonumber\\
&\leq\!\!\!\!\!\!\!\!\sup_{m\in\left\{1,\cdots,\mathcal{M}_{2\epsilon}\paren{\mathcal{S}}\right\}}\!\!\!\!\!\!\!\PRP{\norm{\vec{x}^\star_{ m}-\mathcal{A}_{T+1}\paren{{X}_{1:T},{\hat{Y}_{1:T}},\hat{Z}_{1:T}}}{}\geq \epsilon}\nonumber
\end{align}
which completes the proof.
\end{IEEEproof}
We next use Fano inequality to obtain a lower bound on $\PRP{E_{\mathcal{A}}}$. To this end, let the random variable $M\in\left\{1,\cdots,\mathcal{M}_{2\epsilon}\paren{\mathcal{S}}\right\}$ encode the choice of utility function vector from the set $\mathcal{F}^\prime$. Also, let the random variable  $\hat{M}\in\left\{1,\cdots,\mathcal{M}_{2\epsilon}\paren{\mathcal{S}}\right\}$ encode the estimated NE by genie. Then, using Fano inequality \cite{CT06}, we have 
\begin{align}\label{Eq: Fano}
\PRP{E_\mathcal{A}}&\geq \frac{\CDSE{M}{\hat{M}}-1}{\log \mathcal{M}_{2\epsilon}\paren{\mathcal{S} }}\nonumber\\
&\stackrel{(a)}{=}\geq 1-\frac{1+\DSE{M}-\CDSE{M}{\hat{M}}}{\log \mathcal{M}_{2\epsilon}\paren{\mathcal{S} }}\nonumber\\
 &=1-\frac{1+\MI{M}{\hat{M}}}{\log\mathcal{M}_{2\epsilon}\paren{\mathcal{S}}}
\end{align}
%\begin{eqnarray}
%\PRP{E_\mathcal{A}}\geq 1-\frac{1+\MI{M}{\hat{M}}}{\log\abs{\mathcal{F}^\prime}}\nonumber
%\end{eqnarray}
where $(a)$ follow from the fact that $\DSE{M}=\log\mathcal{M}_{2\epsilon}\paren{\mathcal{S}}$ since $M$ is uniformly distributed over $\left\{1,\cdots,\mathcal{M}_{2\epsilon}\paren{\mathcal{S}}\right\}$. Using \eqref{Eq: Subset-Bound}, \eqref{Eq: Fano} and Lemma \ref{Lem: HT}, we have 
\begin{align}\label{Eq: E-1}
\delta\geq 1-\frac{1+\MI{M}{\hat{M}}}{\log\mathcal{M}_{2\epsilon}\paren{\mathcal{S}}}
\end{align}
Next, we obtain an upper bound on $\MI{M}{\hat{M}}$ using information theoretic inequalities. 
\subsubsection{Applying information theoretic inequalities}
First note that $\paren{M, \left\{X_{1:T},\hat{Y}_{1:T},\hat{Z}_{1:T}\right\},\hat{M}}$ form a Markov chain as follows: $M\longrightarrow X_{1:T},\hat{Y}_{1:T},\hat{Z}_{1:T}\longrightarrow\hat{M}$. Therefore, we have 
\begin{align}\label{Eq: E0}
\MI{M}{\hat{M}}\leq \MI{M}{X_{1:T},\hat{Y}_{1:T},\hat{Z}_{1:T}}.
\end{align}
 Using the chain rule for mutual information, $\MI{M}{X_{1:T},\hat{Y}_{1:T},\hat{Z}_{1:T}}$ can be expanded as 
\begin{align}\label{Eq: main-expansion}
\MI{M}{X_{1:T},\hat{Y}_{1:T},\hat{Z}_{1:T}}&=\sum_{k=1}^T\CMI{M}{X_k,\hat{Y}_k,\hat{Z}_{k}}{X_{1:k-1},\hat{Y}_{1:k-1},\hat{Z}_{1:k-1}}
\end{align}
where $X_k=\left[x^i_k\right]_i$ is the collection of players' actions at time $k$, $\hat{Y}_k=\left[\hat{y}^i_{k}\right]_i$ is the collection of all received utility-related information by players at time $k$, and $\hat{Z}_k=\left\{\hat{z}^{i,p}_{k}:p\in\mathcal{L}_i\right\}_{i}$ is the collection of all constraint-related information received by players at time $k$ where $\mathcal{L}_{i}$ is the set of constraints affected by the $i$th player's action. 

Using the chain rule for conditional mutual information, we have %We can expand  $\CMI{M}{X_i,\hat{Y}_i,\hat{Z}_i}{X_{1:i-1},\hat{Y}_{1:i-1},\hat{Z}_{1:i-1}}$ as 
\begin{align}
\hspace{-1cm}\CMI{M}{X_k,\hat{Y}_k,\hat{Z}_{k}}{X_{1:k-1},\hat{Y}_{1:k-1},\hat{Z}_{1:k-1}}&=\CMI{M}{{X}_k}{X_{1:k-1},\hat{Y}_{1:k-1},\hat{Z}_{1:k-1}}+\CMI{M}{\hat{Z}_k}{X_{1:k},\hat{Y}_{1:k-1},\hat{Z}_{1:k-1}}\nonumber\\
&+\CMI{M}{\hat{Y}_k}{X_{1:k},\hat{Y}_{1:k-1},\hat{Z}_{1:k}}\nonumber
%&\stackrel{(a)}{=}\CMI{M}{\hat{Y}_k}{X_{1:k},\hat{Y}_{1:k-1},\hat{Z}_{1:k}}\nonumber
\end{align}
%where $(a)$ follows from the facts that and \linebreak $\paren{M,X_{1:k-1},\hat{Y}_{1:k-1},\hat{Z}_{1:k-1}}\longrightarrow X_k\longrightarrow\hat{Z}_k$. 
Note that $\hat{Z}_k$ can be written as $\hat{Z}_k=\left\{g_p\paren{\hat{X}_k^{{\rm csn}_{\phi\paren{p}}}}+V^{i,p}_{k}:p\in\mathcal{L}_i\right\}_{i}$ where $\hat{X}_k^{{\rm csn}_{\phi\paren{p}}}=\left\{x^i_k+W^{i}_{k,{\rm csn}_{\phi\paren{p}}}\right\}_{i\in\mathcal{N}_{{\rm csn}_{\phi\paren{p}}}}$. Thus, given $X_k$, $\hat{Z}_k$ only depends on $\left\{W^{i}_{k,{\rm csn}_{\phi\paren{p}}}:i\in\mathcal{N}_{{\rm csn}_{\phi\paren{p}}}\right\}_p$ and $\left\{V^{i,p}_{k}:p\in\mathcal{L}_i\right\}_i$ which are independent of \linebreak $\left\{M,X_{1:k-1},\hat{Y}_{1:k-1},\hat{Z}_{1:k-1}\right\}$. Thus, we have $M,X_{1:k-1},\hat{Y}_{1:k-1},\hat{Z}_{1:k-1}\longrightarrow X_k\longrightarrow\hat{Z}_k$ and 
\begin{align}
&\CMI{M}{\hat{Z}_k}{X_{1:k},\hat{Y}_{1:k-1},\hat{Z}_{1:k-1}}=0\nonumber
\end{align}
Also, we have 
\begin{align}
&\CMI{M}{{X}_k}{X_{1:k-1},\hat{Y}_{1:k-1},\hat{Z}_{1:k-1}}=0\nonumber
\end{align}
 since $x^i_k=\mathcal{A}^i_k\paren{X^i_{1:k-1},\hat{Y}^i_{1:k-1},\hat{Z}^i_{1:k-1}}$, and the collection of random variables $\left(M,\left\{X_{1:k-1},\hat{Y}_{1:k-1},\hat{Z}_{1:k-1}\right\},X_k\right)$ from a Markov chain as follows $M\longrightarrow X_{1:k-1},\hat{Y}_{1:k-1},\hat{Z}_{1:k-1}\longrightarrow{X}_k$. Thus,  we have 
\begin{eqnarray}\label{Eq: E1}
\MI{M}{X_{1:T},\hat{Y}_{1:T},\hat{Z}_{1:T}}=\sum_{k=1}^T\CMI{M}{\hat{Y}_k}{X_{1:k},\hat{Y}_{1:k-1},\hat{Z}_{1:k}}
\end{eqnarray}

Now, $\CMI{M}{\hat{Y}_k}{X_{1:k},\hat{Y}_{1:k-1},\hat{Z}_{1:k}}$ can be upper bounded as 
\begin{align}\label{Eq: E2}
\hspace{-1cm}\CMI{M}{\hat{Y}_k}{X_{1:k},\hat{Y}_{1:k-1},\hat{Z}_{1:k}}&\leq \CMI{M,Y_k}{\hat{Y}_k}{X_{1:k},\hat{Y}_{1:k-1},\hat{Z}_{1:k}}\nonumber\\
&=\CMI{Y_k}{\hat{Y}_k}{\!X_{1:k},\!\hat{Y}_{1:k-1},\!\hat{Z}_{1:k}}+\CMI{M}{\hat{Y}_k}{X_{1:k},\!\hat{Y}_{1:k-1},\!Y_k,\!\hat{Z}_{1:k}}
\end{align}
where $Y_k=\left[y_{k}^i\right]_i$ is the collection of utility-related information computed by the USNs at time $k$.
Using the definition of conditional mutual information, we have 
\begin{align}
\hspace{-1cm}\CMI{M}{\hat{Y}_k}{X_{1:k},\hat{Y}_{1:k-1},Y_k,\hat{Z}_{1:k}}&=\CDE{\hat{Y}_k}{X_{1:k},\!\hat{Y}_{1:k-1},\!Y_k,\!\hat{Z}_{1:k}\!}
-\!\CDE{\hat{Y}_k}{X_{1:k},\!\hat{Y}_{1:k-1},\!Y_k,\!\hat{Z}_{1:k},M\!}\nonumber
\end{align}
 Note that $\hat{Y}_k$ can be written as $\hat{Y}_k=\left[y^i_{k}+V^i_k\right]_i$. Thus, given $Y_k$, $\hat{Y}_k$ only depends on $\left\{V^i_k\right\}_i$ which   is independent of $\left\{X_{1:k},\hat{Y}_{1:k-1},Y_k,\hat{Z}_{1:k},M\right\}$. Thus, random variables $\paren{\left\{M,X_{1:k},\hat{Y}_{1:k-1},\hat{Z}_{1:k}\right\},Y_k,\hat{Y}_k}$ form a Markov chain as \linebreak $M,X_{1:k},\hat{Y}_{1:k-1},\hat{Z}_{1:k}\longrightarrow Y_k\longrightarrow\hat{Y}_k$. Hence, we have  $\CDE{\hat{Y}_k}{M,X_{1:k},\hat{Y}_{1:k-1},Y_k,\hat{Z}_{1:k}}=\CDE{\hat{Y}_k}{Y_k}$. It is straight forward to verify that random variables $\paren{\left\{X_{1:k},\hat{Y}_{1:k-1},\hat{Z}_{1:k}\right\},Y_k,\hat{Y}_k}$ form a Markov chain as $X_{1:k},\hat{Y}_{1:k-1},\hat{Z}_{1:k}\longrightarrow Y_k\longrightarrow\hat{Y}_k$, thus $\CDE{\hat{Y}_k}{X_{1:k},\hat{Y}_{1:k-1},Y_k,\hat{Z}_{1:k}}=\CDE{\hat{Y}_k}{Y_k}$ which implies 
\begin{align}\label{Eq: E3}
&\CMI{M}{\hat{Y}_k}{X_{1:k},\hat{Y}_{1:k-1},Y_k,\hat{Z}_{1:k}}=0
\end{align}
Now, we expand $\CMI{Y_k}{\hat{Y}_k}{X_{1:k},\hat{Y}_{1:k-1},\hat{Z}_{1:k}}$ as follows 
\begin{align}\label{Eq: E4}
\CMI{Y_k}{\hat{Y}_k}{X_{1:k},\hat{Y}_{1:k-1},\hat{Z}_{1:k}}&=\CDE{\hat{Y}_k}{X_{1:k},\hat{Y}_{1:k-1},\hat{Z}_{1:k}}-\CDE{\hat{Y}_k}{Y_k,X_{1:k},\hat{Y}_{1:k-1},\hat{Z}_{1:k}}\nonumber\\
&=\CDE{\hat{Y}_k}{X_{1:k},\hat{Y}_{1:k-1},\hat{Z}_{1:k}}-\CDE{\hat{Y}_k}{Y_k}\nonumber\\
&\stackrel{(a)}{\leq} \DE{\hat{Y}_k}-\CDE{\hat{Y}_k}{Y_k}\nonumber\\
&=\MI{Y_k}{\hat{Y}_k}
\end{align}
where $(a)$ follows from the fact that conditioning reduces entropy. Combining \eqref{Eq: E1}-\eqref{Eq: E4}, we have 
\begin{align}\label{Eq: E5}
\MI{M}{X_{1:T},\hat{Y}_{1:T},\hat{Z}_{1:T}}&\leq \sum_{k=1}^T\MI{Y_k}{\hat{Y}_k}\nonumber\\
&\leq T\max_{p_{Y}\paren{y},\ES{\norm{Y}{}^2}\leq \alpha}\MI{Y}{\hat{Y}}\nonumber\\
&=T C_{\rm down}
\end{align}
%where $C_{\rm down}=\max_{p_{Y}\paren{y},\ES{\norm{Y}{}^2}\leq \alpha}\MI{Y}{\hat{Y}}$ is the total capacity of downlink communication channels from USNs to players, $Y=\left[y^i\right]_i$, $\hat{Y}=\left[\hat{y}^i\right]_i$, $y^i$ and $\hat{y}^i$ are the input  and output of the downlink channel from $\text{USN}_{\pi\paren{i}}$ to the $i$th player, respectively, $p_{Y}\paren{y}$ is the joint distribution of $Y$, and $\alpha$ is the total  transmission power constraint on these channels. 

 Combining  \eqref{Eq: E-1}, \eqref{Eq: E0} and \eqref{Eq: E5}, we have 
\begin{eqnarray}
T\geq \frac{\paren{1-\delta}\log\mathcal{M}_{2\epsilon}\paren{\mathcal{S}}-1}{C_{\rm down}}\nonumber
\end{eqnarray}
which completes the proof.
%%%%%%%%%%%%%%%%%%%%%%%%%%%%%%%%%%%%%%%%%%%%%%%%%%%%%%%%%%%%%%%
\subsection{Proof of Corollary \ref{Cor: DownCap}}\label{Proof: Cor-DownCap}
Let $D$ be a diagonal matrix with diagonal entries equal to $2\epsilon$. Let $\mathcal{D}$ be the lattice $D\Z^N$, \emph{i.e.,} $\mathcal{D}=\left\{D\vec{z}, \vec{z}\in\Z^N\right\}$. Note that the elements of $\mathcal{D}$ are at least $2\epsilon$ apart. Let $\abs{\mathcal{D}\cap\mathcal{S}}$ be the number of  lattice points of $\mathcal{D}$ which lie in $\mathcal{S}$. Clearly, $\mathcal{M}_{2\epsilon}\paren{\mathcal{S}}$ is lower bounded by $\abs{\mathcal{D}\cap\mathcal{S}}$. We use the following result from \cite{S93} to obtain a lower bound on $\abs{\mathcal{D}\cap\mathcal{S}}$ in terms of $\epsilon$, volume and surface area of $\mathcal{S}$.  
\begin{lemma}{\cite{S93}}
Let $\mathcal{D}$ be a lattice in $\R^N$ with non-zero determinant, \emph{i.e.,} ${\rm Det}\paren{D}\neq 0$. Let $\mathcal{S}$ be a convex and compact subset of $\R^N$. Then, we have 
\begin{eqnarray}
\abs{\mathcal{D}\cap\mathcal{S}}\geq \frac{1}{{\rm Det}\paren{\mathcal{D}}}\paren{{\rm Vol}\paren{\mathcal{S}}-\frac{\lambda_N\paren{\mathcal{D}}}{2}{\rm P}\paren{\mathcal{S}}}
\end{eqnarray}
where ${\rm Vol}\paren{\mathcal{S}}$ is the volume of $\mathcal{S}$, ${\rm P}\paren{\mathcal{S}}$ is the surface area of $\mathcal{S}$ and $\lambda_N\paren{\mathcal{D}}$ is the successive minima of $\mathcal{D}$ defined as the smallest $\rho$ such that there exist $N$ linearly independent elements of lattice, $\left\{d_1,\cdots,d_N\in\mathcal{D}\backslash \left\{\vec{0}\right\}\right\}$ such that $\norm{d_i}{}\leq \rho$ \cite{FS99}. 
\end{lemma} 
For the lattice $\mathcal{D}=D\Z^N$, we have ${\rm Det}\paren{\mathcal D}=\paren{2\epsilon}^N$ and $\lambda_N\paren{\mathcal{D}}=2\epsilon$. Thus, $\mathcal{M}_{2\epsilon}\paren{\mathcal{S}}$ can be lower bounded as 
\begin{eqnarray}
\mathcal{M}_{2\epsilon}\paren{\mathcal{S}}\geq \paren{\frac{1}{2\epsilon}}^N\paren{{\rm Vol}\paren{\mathcal{S}}-\epsilon{\rm P}\paren{\mathcal{S}}}
\end{eqnarray}
%%%%%%%%%%%%%%%%%%%%%%%%%%%%%%%%%%%%%%%%%%%%%%%%%%%%%%%%%%%%%%%%%%%%%%%%%%%%%%%%%%%%%%%%%%%%%%
\subsection{Proof of Theorem \ref{Theo: Gaussian-bound}}\label{Proof:  Gaussian-bound}
Similar to the proof of Theorem \ref{Theo: DownCap}, we first construct a finite subset of $\mathcal{F}_\gamma$ as follows. Recall that $\mathcal{S}$ contains a 2-ball of radius $\sqrt{2}\epsilon$. Let ${\rm B}_{\sqrt{2}\epsilon}$ denote such a ball. Also, let 
\begin{eqnarray}
\mathcal{S}^\star_{2\epsilon}=\left\{\vec{x}^\star_1,\cdots,\vec{x}^\star_4\right\}
\end{eqnarray}
be the set of four points in  ${\rm B}_{\sqrt{2}\epsilon}$ which are 90 degrees apart. Thus, we have $\max_{\vec{x}^\star_m,\vec{x}^\star_{m^\prime}\in\mathcal{S}^\star_{2\epsilon}}\norm{\paren{\vec{x}^\star_m-\vec{x}^\star_{m^\prime}}}{}=2\sqrt{2}\epsilon$ and $\abs{\mathcal{S}^\star_{2\epsilon}}=4$. 

Here, for each $\vec{x}^\star_m\in\mathcal{S}^\star_{2\epsilon}$ $\left(m=1,\cdots,4\right)$, we construct a utility function vector $U_m\paren{\vec{x}}$ such that $\vec{x}^\star_m$ becomes the Nash equilibrium (NE) of the non-cooperative game with $N$ players, utility function vector $U_m\paren{\vec{x}}$ and the action space $\mathcal{S}$. To this end, let $A=\left[a_{ij}\right]_{i,j}$ be an $N$-by-$N$, symmetric, negative definite matrix with $\norm{A}{}=\gamma$. Then, the utility function vector $U_m\paren{\vec{x}}$ is constructed as  $U_m\paren{\vec{x}}=\left[u_{m,i}\paren{x^i,\vec{x}^{-i}}\right]_i$ where $u_{m,i}\paren{x^i,\vec{x}^{-i}} =\frac{a_{ii}}{2}\paren{x^i}^2+x^i\paren{-A_i\vec{x}^\star_m+\sum_{j\neq i} a_{ij}x^j}$ and $A_i$ is the $i$th row of $A$.  It is straight forward to verify that $\vec{x}^\star_{m}$ is a NE of the $N$-player non-cooperative game with the utility function vector $U_{m}\paren{\vec{x}}$ and the constraint set $\mathcal{S}$ (see the proof of Lemma \ref{Lem: Uniq} in Subsection \ref{Proof: DownCap}). Let 
\begin{eqnarray}
\mathcal{F}^\prime_\gamma=\left\{U_{m}\paren{\vec{x}},m=1,\cdots,4\right\}
\end{eqnarray}
 denote a finite set of utility vector functions. 
 Since we have $J_{\tilde{\nabla} U_m\paren{\vec{x}}}=A$ for $m=1,\cdots,4$, each utility function vector $U_m\paren{\cdot}$ belongs to the function class $\mathcal{F}_{\gamma}$. Hence, $\mathcal{F}^\prime_\gamma$ is a subset of $\mathcal{F}_\gamma$. The class of $N$-player non-cooperative games with utility function vectors in $\mathcal{F}^\prime_\gamma$  and the action space $\mathcal{S}$  is denoted as $\mathcal{G}^\prime_\gamma=\langle \mathcal{N}, \mathcal{S},\mathcal{F}^\prime_\gamma\rangle$. Here, we make the technical assumption that each game in $\mathcal{G}^\prime_\gamma$ admits a unique NE. 

%  such that $(i)$ each game in the game class $\mathcal{G}^\prime_\gamma=\langle \mathcal{N}, \mathcal{S},\mathcal{F}^\prime\rangle$ admits a unique Nash equilibrium, $(ii)$ $\norm{J_{F_{U_m}}\paren{\vec{x}}}{}=\gamma$ {\bf change according to the text}. Let $\mathcal{S}^\star_{2\epsilon}=\left\{\vec{x}_{m}\in\mathcal{S},m=1,\cdots,\abs{\mathcal{F}^\prime_\gamma}\right\}$ denote a $2\epsilon$-distinguishable subset of $\mathcal{S}$. The utility functions $U_{m}\paren{\vec{x}}$s are selected such that the set $\mathcal{S}^\star_{2\epsilon}$ coincides with the set of all Nash equilibria due to the games in $\mathcal{G}^\prime_\gamma$. Note that, different from Subsection \ref{App: DownCap}, the set $\mathcal{S}^\star_{2\epsilon}$ is a $2\epsilon$-distinguishable subset of $\mathcal{S}$ but not necessarily a maximal size one. The set $\mathcal{F}^\prime_\gamma$ is constructed as follows.
 
For a given $\epsilon$ and $\delta<\frac{1}{2}$,  consider any algorithm $\mathcal{A}$ such that after $T$ time-steps, we have 
\begin{align}
\!\!\!\!\sup_{U\paren{\cdot}\in \mathcal{F}_\gamma}\!\!\inf_i\PRP{\norm{\vec{x}_{{\rm NE}_i, U\paren{\cdot}}\!\!-\!\!\mathcal{A}_{T+1}\!\!\paren{{X}_{1:T},{\hat{Y}_{1:T}},\hat{Z}_{1:T}}}{}\geq \epsilon}\leq \delta.\nonumber
\end{align}
 Since $\mathcal{F}^\prime_\gamma$ is a subset of $\mathcal{F}_\gamma$ and the games in $\mathcal{G}^\prime_\gamma$ admit a unique NE, we have 
\begin{align}
\sup_{m=1,\cdots,\abs{\mathcal{S}^\star_{2\epsilon}}}\PRP{\norm{\vec{x}^\star_{m}-\mathcal{A}_{T+1}\paren{{X}_{1:T},{\hat{Y}_{1:T}},\hat{Z}_{1:T}}}{}\geq \epsilon}\leq \delta.\nonumber
\end{align}
Consider a genie-aided hypothesis test in which a genie  selects a game instance from $\mathcal{G}^\prime_\gamma$ uniformly at random. Let $\vec{x}^\star_M\in\mathcal{S}^\star_{2\epsilon}$ and  $U_M\paren{\cdot}\in\mathcal{F}^\prime$ denote the NE and the  utility function vector associated with the randomly selected game instance, respectively, where $M$ is a random variable uniformly distributed over the set $\left\{1,\cdots,4\right\}$. Also, let the random variable $\hat{M}=\left\{1,\cdots,4\right\}$ encode the outcome of the genie-aided hypothesis test in Subsection \ref{Proof: DownCap}. Then, using Lemma \ref{Lem: HT}, Fano inequality and the fact that $\abs{\mathcal{S}^\star_{2\epsilon}}=4$, we have 
\begin{align}\label{Eq: C-E0}
\delta\geq 1-\frac{1+\MI{M}{\hat{M}}}{2}
\end{align}
 Using \eqref{Eq: E1} in Subsection  \ref{Proof: DownCap}, we have 
\begin{eqnarray}\label{Eq: Aux-B-1}
\MI{M}{\hat{M}}\leq \sum_{k=1}^T\CMI{M}{\hat{Y}_k}{X_{1:k},\hat{Y}_{1:k-1},\hat{Z}_{1:k}}
\end{eqnarray}
Under the partial-derivative computation model for USNs, $y^i_{k}$ can be written as
\begin{align}
y^i_{k}&=-A_i\vec{x}^\star_M+\sum_{j\in\mathcal{N}_{{\rm usn}_{\pi\paren{i}}}} a_{ij}\hat{x}^{j}_{k,{\rm usn}_{\pi\paren{i}}}\nonumber\\
&=-A_i\vec{x}^\star_M+\sum_{j\in\mathcal{N}_{{\rm usn}_{\pi\paren{i}}}} a_{ij}\paren{x^j_k+W^{j}_{k,{\rm usn}_{\pi\paren{i}}}}\nonumber
\end{align}
Thus, $\hat{y}^i_{k}$ can be written as
\begin{align}
\hat{y}^i_{k}&=-A_i\vec{x}^\star_M+\paren{\sum_{j\in\mathcal{N}_{{\rm usn}_{\pi\paren{i}}}} a_{ij}\paren{x^j_k+W^{j}_{k,{\rm usn}_{\pi\paren{i}}}}}+V^i_{k}\nonumber\\
&=-A_i\vec{x}^\star_M+\paren{\sum_{j} a_{ij}x^j_k}+\hat{W}_k^i\nonumber
\end{align}
where $\hat{W}^i_k=\paren{\sum_{j\in\mathcal{N}_{{\rm usn}_{\pi\paren{i}}}}a_{ij}W^{j}_{k,{\rm usn}_{\pi\paren{i}}}}+V^i_{k}$. Note that $\hat{Y}_k=\left[\hat{y}^i_{k}\right]_i$ can be written as $\hat{Y}_k=AX_k-A\vec{x}^\star_M+\hat{W}_k$ where $X_k=\left[x^i_k\right]_i$ and $\hat{W}_k=\left[\hat{W}_k^i\right]_i$.%% The random vector  $\hat{W}_k$ depends on the communication noise between players and USNs at time $k$ and is independent of $\left\{M,X_{1:k},\hat{Y}_{1:k-1},\hat{Z}_{1:k}\right\}$. 

Note that $\CMI{M}{\hat{Y}_k}{X_{1:k},\hat{Y}_{1:k-1},\hat{Z}_{1:k}}$ can be upper bounded as:
\begin{align}\label{Eq: Aux-B-2}
\CMI{M}{\hat{Y}_k}{X_{1:k},\hat{Y}_{1:k-1},\hat{Z}_{1:k}}&\stackrel{(a)}{=}\CDE{\hat{Y}_k}{X_{1:k},\hat{Y}_{1:k-1},\hat{Z}_{1:k}}-\CDE{\hat{Y}_k}{M,X_{1:k},\hat{Y}_{1:k-1},\hat{Z}_{1:k}}\nonumber\\
&=\CDE{AX_k-A\vec{x}^\star_M+\hat{W}_k}{X_{1:k},\hat{Y}_{1:k-1},\hat{Z}_{1:k}}\nonumber\\
&-\CDE{AX_k-A\vec{x}^\star_M+\hat{W}_k}{M,X_{1:k},\hat{Y}_{1:k-1},\hat{Z}_{1:k}}\nonumber\\
&\stackrel{(b)}{=}\CDE{-A\vec{x}^\star_M+\hat{W}_k}{X_{1:k},\hat{Y}_{1:k-1},\hat{Z}_{1:k}}\nonumber\\
&\hspace{2cm}-\CDE{\hat{W}_k}{M,X_{1:k},\hat{Y}_{1:k-1},\hat{Z}_{1:k}}\nonumber\\
&\stackrel{(c)}{=}\CDE{-A\vec{x}^\star_M+\hat{W}_k}{X_{1:k},\hat{Y}_{1:k-1},\hat{Z}_{1:k}}-\DE{\hat{W}_k}\nonumber\\
&\stackrel{(d)}{\leq} \DE{-A\vec{x}^\star_M+\hat{W}_k}-\DE{\hat{W}_k}\nonumber\\
&\stackrel{(e)}{=} \DE{- A\vec{x}^\star_M+\hat{W}_k}-\CDE{-A\vec{x}^\star_M+\hat{W}_k}{\vec{x}^\star_M}\nonumber\\
&\stackrel{(e)}{=} \DE{- A\vec{x}^\star_M+\hat{W}_1}-\CDE{-A\vec{x}^\star_M+\hat{W}_1}{\vec{x}^\star_M}\nonumber\\
&=\MI{\vec{x}^\star_M}{-A\vec{x}^\star_M+\hat{W}_A}
%&\stackrel{(d)}{=} \DE{AN^{{\rm up},y}_k+b_M+N^{{\rm down},y}_k}\nonumber\\
%&\hspace{2cm}-\CDE{AN^{{\rm up},y}_k+N^{{\rm down},y}_k+ b_M}{b_M}\nonumber\\
%&\stackrel{(e)}=\MI{b_M}{AN^{{\rm up},y}_k+N^{{\rm down},y}_k+ b_M}\nonumber
%&=\DSE{b_M}-\CDSE{b_M}{AN^{{\rm up},y}_i+N^{{\rm down},y}_i+ b_M}\nonumber\\
%&\stackrel{(f)}{\leq} \DSE{b_M} +\logp{1-\mathsf {Pe}\paren{b_M}}\nonumber\\
%&\stackrel{(g)}{=}\log\mathcal{M}_{2\epsilon}\paren{\mathcal{S}}+\logp{1-\mathsf{Pe}\paren{b_M}}
\end{align}
where $\hat{W}_A=\left[\hat{W}^i_1\right]_i$ with $\hat{W}^i_1=\sum_{j\in\mathcal{N}_{{\rm usn}_{\pi\paren{i}}}}\paren{a_{ij}W^{j}_{1,{\rm usn}_{\pi\paren{i}}}}+V^i_{1}$,  $(a)$ follows from the definition of conditional mutual information, $(b)$ follows from the translation invariance property of differential entropy, $(c)$ follows from the fact that $\hat{W}_k$ is independent of $\left\{M,X_{1:k},\hat{Y}_{1:k-1},\hat{Z}_{1:k}\right\}$, $(d)$ follows from the fact that conditioning reduces entropy, and $(e)$ follows from the translation invariance property of the differential entropy and the fact that the random vectors $\hat{W}_1$ and $\hat{W}_k$ have the same probability density functions (PDFs). Combining \eqref{Eq: Aux-B-1} and \eqref{Eq: Aux-B-2}, we have 
\begin{eqnarray}\label{Eq: C-E2}
\MI{M}{\hat{M}}\leq T \MI{\vec{x}^\star_M}{-A\vec{x}^\star_M+\hat{W}_A}
\end{eqnarray}

Using the convexity of the Kullback-Leibler (KL) distance, $\MI{\vec{x}^\star_M}{-A\vec{x}^\star_M+\hat{W}_A}$ can be upper bounded as \eqref{Eq: KL}
\begin{figure*}
\begin{align}\label{Eq: KL}
\MI{\vec{x}^\star_M}{-A\vec{x}^\star_M+\hat{W}_A}&=\ESI{\vec{x}^\star_M}{\KLD{\CD{-A\vec{x}^\star_M+\hat{W}_A}{\vec{x}^\star_M}}{p_{-A\vec{x}^\star_M+\hat{W}_A}\paren{\vec{x}}}}\nonumber\\
&\stackrel{(a)}{=}\ESI{\vec{x}^\star_M}{\KLD{\ESI{\vec{x}^\star_{M^\prime}}{\CD{-A\vec{x}^\star_M+\hat{W}_A}{\vec{x}_M}}}{\ESI{\vec{x}^\star_{M^\prime}}{\CD{-A\vec{x}^\star_{M^\prime}+\hat{W}_A}{\vec{x}^\star_{M^\prime}}}}}\nonumber\\
&\stackrel{(b)}{\leq} \ESI{\vec{x}^\star_M}{\ESI{\vec{x}^\star_{M^\prime}}{\KLD{\CD{-A\vec{x}^\star_M+\hat{W}_A}{\vec{x}^\star_M}}{\CD{-A\vec{x}^\star_{M^\prime}+\hat{W}_A}{\vec{x}^\star_{M^\prime}}}}}\nonumber\\
%&= \frac{1}{\abs{\mathcal{S}^\star_{2\epsilon}}^2}\sum_{\vec{x}_M,\vec{x}_{M^\prime}}\KLD{\CD{-A\vec{x}_M+\hat{W}_A}{\vec{x}_M}}{\CD{-A\vec{x}_{M^\prime}+\hat{W}_A}{\vec{x}_{M^\prime}}}\nonumber\\
&\leq \max_{\vec{x}^\star_m,\vec{x}^\star_{m^\prime}\in\mathcal{S}^\star_{2\epsilon}}\KLD{p_{-A\vec{x}^\star_m+\hat{W}_A}\paren{\vec{x}}}{p_{-A\vec{x}^\star_{m^\prime}+\hat{W}_A}\paren{\vec{x}}}
\end{align}
\hrule
\end{figure*}
where $\KLD{p\paren{x}}{q\paren{x}}$ is the KL distance between the pair of PDFs $\paren{p\paren{x},q\paren{x}}$, $\vec{x}^\star_{M^\prime}$ is a random vector taking value in $\mathcal{S}^\star_{2\epsilon}$ with uniform distribution, independent of $\vec{x}^\star_M$, $(a)$ follows from the fact that  
\begin{align}
p_{-A\vec{x}^\star_M+\hat{W}_A}\paren{\vec{x}}=\ESI{\vec{x}^\star_{M^\prime}}{\CD{-A\vec{x}^\star_{M^\prime}+\hat{W}_A}{\vec{x}^\star_{M^\prime}}}\nonumber
\end{align}
% where $\CD{-A\vec{x}^\star_{M^\prime}+\hat{W}_A}{\vec{x}^\star_{M^\prime}}$ is the conditional PDF of $-A\vec{x}^\star_{M^\prime}+\hat{W}_A$ given $\vec{x}^\star_{M^\prime}$, 
  since the random vectors $-A\vec{x}^\star_M+\hat{W}_A$ and $-A\vec{x}^\star_{M^\prime}+\hat{W}_A$ have the same joint PDFs, and $(b)$ follows from the convexity of $\KLD{p\paren{x}}{q\paren{x}}$  in  $\paren{p\paren{x},q\paren{x}}$. 

To evaluate the KL term in \eqref{Eq: KL}, we need to study the joint PDF of the random vector $-A\vec{x}^\star_m+\hat{W}_A$. Note that random vector $-A\vec{x}^\star_m+\hat{W}_A$ is a Gaussian distributed random vector with mean $-A\vec{x}^\star_m$. The next lemma provides an expression for the covariance matrix of $A\vec{x}^\star_m+\hat{W}_A$.  
\begin{lemma}
Let $\Sigma_A$ be an $N$-by-$N$ matrix defined as 
\begin{align}
\Sigma_A=&\diag{\sigma^2_1,\cdots,\sigma^2_N}+\diag{\sigma^2_{{\rm usn}_{\pi\paren{1}}},\cdots,\sigma^2_{{\rm usn}_{\pi\paren{N}}}}A A\nonumber 
\end{align} 
Also, let $G=\left[G_{ij}\right]_{ij}$ denote an $N$-by-$N$ matrix defined as 
\begin{align}
G_{ij}=\left\{
\!\!\!\!\begin{array}{cc}
1&{\rm if}\quad \pi\paren{i}=\pi\paren{j}  \nonumber\\
0&\quad {\rm Otherwise} \nonumber\\
%\sigma^2_i+\sum_ta^2_{it}\sigma^2_{l_i,{\rm usn}}&{\rm if}\quad i=j \nonumber\\
\end{array}
\right.
\end{align}
Then, the covariance matrix of $-A\vec{x}^\star_m+\hat{W}_A$ can be written as $\Sigma_A\circ G$ where $\circ$ represents the Hadamard product.
\end{lemma}
\begin{IEEEproof}
Note that the covariance matrix of $-A\vec{x}^\star_m+\hat{W}_A$ is the same as that of $\hat{W}_A=\left[\hat{W}^i_1\right]_i$ where \linebreak $\hat{W}^i_1=\sum_{j\in\mathcal{N}_{{\rm usn}_{\pi\paren{i}}}}\paren{a_{ij}W^{j}_{1,{\rm usn}_{\pi\paren{i}}}}+V^i_{1}$. The covariance of the $i$th and $t$th entries of $\hat{W}_A$ can be written as 
\begin{align}
\ES{\hat{W}^i_1\hat{W}^t_1}\!\!=\!\!\left\{
\!\!\!\!\begin{array}{cc}
\sigma^2_i\delta\left[i-t\right]\!+\!\sigma^2_{{\rm usn}_{\pi\paren{i}}}\!\!A_i\paren{A_t}^\top&{\rm if}\quad \pi\paren{i}=\pi\paren{t}  \nonumber\\
0&\quad {\rm Otherwise} \nonumber\\
%\sigma^2_i+\sum_ta^2_{it}\sigma^2_{l_i,{\rm usn}}&{\rm if}\quad i=j \nonumber\\
\end{array}
\right.
\end{align}
where $\delta\left[\cdot\right]$ denotes the Kronecker delta function, and $A_i$ is the $i$th row of $A$. Using the definition of the matrix $G$, we have 
\begin{align}
\ES{\hat{W}^i_1\hat{W}^t_1}=\sigma^2_i\delta\left[i-t\right] +\!\sigma^2_{{\rm usn}_{\pi\paren{i}}}\!\!A_i\paren{A_t}^\top G_{it}\nonumber
\end{align}
Thus, the covariance of $\hat{W}_A$, \emph{i.e.,} $C_{\hat{W}_A}$, can be written as 
\begin{align}
C_{\hat{W}_A}=&\diag{\sigma^2_1,\cdots,\sigma^2_N}+\paren{\diag{\sigma^2_{{\rm usn}_{\pi\paren{1}}},\cdots,\sigma^2_{{\rm usn}_{\pi\paren{N}}}}A A^\top}\circ G\nonumber\\
\stackrel{(a)}{=}&\left(\diag{\sigma^2_1,\cdots,\sigma^2_N}+\diag{\sigma^2_{{\rm usn}_{\pi\paren{1}}},\cdots,\sigma^2_{{\rm usn}_{\pi\paren{N}}}}A A\right)\circ G\nonumber\\
=&\Sigma_A\circ G\nonumber
\end{align}
where $(a)$ follows from the fact that the matrix $A$ is symmetric.
\end{IEEEproof}
% completes the proof.% Similarly,  $-A\vec{x}_{m^\prime}+\hat{W}_A$ is a Gaussian distributed random vector with mean $-A\vec{x}_{m^\prime}$ and the covariance matrix $\Sigma_A\circ G$. 

To use the expression of KL distance between two Gaussian PDFs, we need to ensure that the matrix $\Sigma_A\circ G$ is invertible. This result is established in the next lemma.
\begin{lemma}\label{Lem: Inv-cov}
The matrix $\Sigma_A\circ G$  is an invertible matrix.
\end{lemma}
\begin{IEEEproof}
 Note that $\Sigma_A$ can be written as 
\begin{align}\label{Eq: Cov}
\Sigma_A\circ G=&\diag{\sigma^2_1,\cdots,\sigma^2_N}+\paren{\diag{\sigma^2_{{\rm usn}_{\pi\paren{1}}},\cdots,\sigma^2_{{\rm usn}_{\pi\paren{N}}}}A A}\circ G
\end{align} 
The second term in \eqref{Eq: Cov} is the covariance of the random vector 
\begin{align}
\left[\sum_{j\in\mathcal{N}_{{\rm usn}_{\pi\paren{i}}}}a_{ij}W^{j}_{1,{\rm usn}_{\pi\paren{i}}}\right]_i,\nonumber
\end{align}
thus, it is a positive semi-definite matrix. Since $\diag{\sigma^2_1,\cdots,\sigma^2_N}$ is positive definite, $\Sigma_A\circ G$ is a positive definite matrix. Hence,  $\Sigma_A\circ G$  is invertible.
\end{IEEEproof}
 Using the expression of KL distance between two Gaussian PDFs, we  have \eqref{Eq: KL-G} where $\norm{A}{}$ and $\norm{\paren{\Sigma_A\circ G}^{-1}}{}$ are the induced matrix norms of $A$ and $\paren{\Sigma_A\circ G}^{-1}$, respectively. 
\begin{figure*}
\begin{align}\label{Eq: KL-G}
\KLD{p_{-A\vec{x}^\star_m+\hat{W}_A}\paren{\vec{x}}}{p_{-A\vec{x}^\star_{m^\prime}+\hat{W}_A}\paren{\vec{x}}}&=\frac{1}{2}\paren{A\paren{\vec{x}^\star_m-\vec{x}^\star_{m^\prime}}}^\top\paren{\Sigma_A\circ G}^{-1}A\paren{\vec{x}^\star_m-\vec{x}^\star_{m^\prime}}\nonumber\\
&\leq \frac{1}{2}\norm{A}{}^2\norm{\paren{\vec{x}^\star_m-\vec{x}^\star_{m^\prime}}}{}^2\norm{\paren{\Sigma_A\circ G}^{-1}}{}\nonumber\\
&=\frac{1}{2}\gamma^2\norm{\paren{\vec{x}^\star_m-\vec{x}^\star_{m^\prime}}}{}^2\norm{\paren{\Sigma_A\circ G}^{-1}}{}
\end{align}
\hrule
\end{figure*}
Recall that the set $\mathcal{S}^\star_{2\epsilon}$ was selected such that $\max_{\vec{x}_m,\vec{x}_{m^\prime}\in\mathcal{S}^\star_{2\epsilon}}\norm{\paren{\vec{x}^\star_m-\vec{x}^\star_{m^\prime}}}{}=2\sqrt{2}\epsilon$ and $\abs{\mathcal{S}^\star_{2\epsilon}}=4$. Using this construction for $\mathcal{S}^\star_{2\epsilon}$, \eqref{Eq: KL} and \eqref{Eq: KL-G}, $\MI{\vec{x}^\star_M}{-A\vec{x}^\star_M+\hat{W}_A}$ can be upper bounded as 
\begin{align}\label{Eq: C-E3}
\MI{\vec{x}^\star_M}{-A\vec{x}^\star_M+\hat{W}_A}\leq 4\gamma^2\epsilon^2\norm{\paren{\Sigma_A\circ G}^{-1}}{}
\end{align}
Note that $\norm{\paren{\Sigma_A\circ G}^{-1}}{}$ can be upper bounded as \eqref{Eq: Norm-upperbound}
\begin{figure*}
\begin{align}\label{Eq: Norm-upperbound}
\norm{\paren{\Sigma_A\circ G}^{-1}}{}&\stackrel{(a)}{=}\lambda_{\rm max}\paren{{\paren{\Sigma_A\circ G}^{-1}}}\nonumber\\
&=\frac{1}{\lambda_{\rm min}\paren{\Sigma_A\circ G}}\nonumber\\
&\stackrel{(b)}{=}\frac{1}{\lambda_{\rm min}\paren{\diag{\sigma^2_1,\cdots,\sigma^2_N}+\diag{\sigma^2_{{\rm usn}_{\pi\paren{1}}},\cdots,\sigma^2_{{\rm usn}_{\pi\paren{N}}}}A A\circ G}}\nonumber\\
&\stackrel{(c)}{\leq}\frac{1}{\lambda_{\rm min}\paren{\diag{\sigma^2_1,\cdots,\sigma^2_N}}+\lambda_{\rm min}\paren{\diag{\sigma^2_{{\rm usn}_{\pi\paren{1}}},\cdots,\sigma^2_{{\rm usn}_{\pi\paren{N}}}}A A\circ G}}\nonumber\\
&\stackrel{(d)}{\leq}\frac{1}{\lambda_{\rm min}\paren{\diag{\sigma^2_1,\cdots,\sigma^2_N}}}\nonumber\\
&=\frac{1}{\min_i \sigma^2_i}
\end{align}
\hrule
\end{figure*}
\textcolor{black}{where $\lambda_{\rm max}\paren{\cdot}$ and $\lambda_{\rm min}\paren{\cdot}$ represent the maximum and minimum eigenvalues, respectively,  $(a)$ follows from the fact that $\paren{\Sigma_A\circ G}^{-1}$ is symmetric and positive definite, $(b)$ follows from \eqref{Eq: Cov}, $(c)$ follows from dual Weyl inequality \cite{Tao12} and $(d)$ follows from the fact that $\diag{\sigma^2_{{\rm usn}_{\pi\paren{1}}},\cdots,\sigma^2_{{\rm usn}_{\pi\paren{N}}}}A A\circ G$ is a positive semi-definite matrix (see the proof of Lemma \ref{Lem: Inv-cov}).} Combining \eqref{Eq: C-E0}, \eqref{Eq: C-E2}, \eqref{Eq: C-E3} and \eqref{Eq: Norm-upperbound}, we have 
\begin{eqnarray}
T\geq \frac{\paren{2\paren{1-\delta}-1}\min_i\sigma^2_i}{4\gamma^2\epsilon^2}\nonumber
\end{eqnarray}
which completes the proof.
%%%%%%%%%%%%%%%%%%%%%%%%%%%%%%%%%%%%%%%%%%%%%%%%%%%%%%%%%%%%%%%%%%%%%%%%%%%%%%%%%%%%%%%%%%%%%%
\subsection{Proof of Theorem \ref{Theo:  Non-Gaussian-bound}}\label{Proof: Gen-noise}
Similar to the proof of Theorem \ref{Theo: Gaussian-bound}, we first restrict our analysis to a finite subset of $\mathcal{F}_\gamma$. To this end, let $\mathcal{S}^\star_{2\epsilon}$ and $\mathcal{F}^\prime_\gamma$ denote the $2\epsilon$-distinguishable subset of $\mathcal{S}$ and the finite subset of $\mathcal{F}_\gamma$, respectively, constructed in the proof of Theorem \ref{Theo:  Gaussian-bound}. For a given $\epsilon$ and $\delta<\frac{1}{2}$,  consider any algorithm $\mathcal{A}$ which can solve any game in $\mathcal{G}_\gamma$ after $T$ time-steps when the uplink channels are noiseless, \emph{i.e.,} 
\begin{align}
\!\!\!\!\sup_{U\paren{\cdot}\in \mathcal{F}_\gamma}\!\!\inf_i\PRP{\norm{\vec{x}_{{\rm NE}_i, U\paren{\cdot}}\!\!-\!\!\mathcal{A}_{T+1}\!\!\paren{{X}_{1:T},{\hat{Y}_{1:T}},\hat{Z}_{1:T}}}{}\geq \epsilon}\leq \delta.\nonumber
\end{align}
 Using the proposed genie-aided hypothesis test in Subsection \ref{Proof: Gaussian-bound}, \eqref{Eq: C-E0} and \eqref{Eq: C-E2}, we have 
\begin{align}\label{Eq: E-E0}
\delta&\geq 1-\frac{1+T \MI{\vec{x}^\star_M}{-A\vec{x}^\star_M+\hat{W}_A}}{2}\nonumber\\
&\stackrel{(a)}{=}1-\frac{1+T \MI{\vec{x}^\star_M}{-A\vec{x}^\star_M+V_1}}{2}
\end{align}
 where $\vec{x}^\star_M$ is a random vector taking value in $\mathcal{S}^\star_{2\epsilon}$ with uniform distribution, $V_1=\left[V_1^1,\cdots,V_1^N\right]^\top$ and $(a)$ follows from the fact that the uplink channels are noiseless. Next, we obtain an upper bound on the mutual information term in \eqref{Eq: E-E0} as follows. In the absence of uplink noise, the inequality \eqref{Eq: KL} can be written as \eqref{Eq: KL-2}
\begin{figure*}
\begin{align}\label{Eq: KL-2}
\MI{\vec{x}^\star_M}{-A\vec{x}^\star_M+V_1}&\leq \max_{\vec{x}^\star_m,\vec{x}^\star_{m^\prime}\in\mathcal{S}^\star_{2\epsilon}}\KLD{p_{-A\vec{x}^\star_m+V_1}\paren{\vec{x}}}{p_{-A\vec{x}^\star_{m^\prime}+V_1}\paren{\vec{x}}}\nonumber\\
&\stackrel{(a)}{=} \max_{\vec{x}^\star_m,\vec{x}^\star_{m^\prime}\in\mathcal{S}^\star_{2\epsilon}}\sum_{i=1}^N\KLD{p_{-A_i\vec{x}^\star_m+V^i_1}\paren{x}}{p_{-A_i\vec{x}^\star_{m^\prime}+V^i_1}\paren{x}}
\end{align}
\hrule
\end{figure*}
where $A_i$ is the $i$th row of matrix $A$ and $(a)$ follows from the fact that the entries of the random vector $V_1$ are jointly independent.

The next lemma derives an asymptotic expansion for \KLD{p_{-A_i\vec{x}^\star_m+V^i_1}\paren{x}}{p_{-A_i\vec{x}^\star_{m^\prime}+V^i_1}\paren{x}}.
\begin{lemma}\label{Lem: KL-expansion}
The KL distance between the probability distribution functions (PDFs) $p_{-A_i\vec{x}^\star_m+V^i_1}\paren{x}$ and $p_{-A_i\vec{x}^\star_{m^\prime}+V^i_1}\paren{x}$ can be written as 
\begin{align}
\KLD{p_{-A_i\vec{x}^\star_m+V^i_1}\paren{x}}{p_{-A_i\vec{x}^\star_{m^\prime}+V^i_1}\paren{x}}=\frac{1}{2}\mathcal{I}_i\paren{ A_i\paren{\vec{x}^\star_{m}-\vec{x}^\star_{m^\prime}}}^2+\BO{\epsilon^3}\nonumber
\end{align}
where $\mathcal{I}_i=-\int_{-\infty}^{\infty}p_{V^i}\paren{x}\frac{d^2}{d x^2} \log p_{V^i}\paren{x} dx$ denotes the Fisher information of $p_{V^i}\paren{x}$  with respect to a shift parameter.
\end{lemma}
\begin{IEEEproof}
To prove this lemma, we first expand $\KLD{p_{-A_i\vec{x}^\star_m+V^i_1}\paren{x}}{p_{-A_i\vec{x}^\star_{m^\prime}+V^i_1}\paren{x}}$ as \eqref{Eq: KL-Down}
\begin{figure*}
\begin{align}\label{Eq: KL-Down}
&\KLD{p_{-A_i\vec{x}^\star_m+V^i_1}\paren{x}}{p_{-A_i\vec{x}^\star_{m^\prime}+V^i_1}\paren{x}}\!\!=\!\!\!\int\!\! p_{V^i}\paren{x+A_i\vec{x}^\star_m}\paren{\log p_{V^i}\paren{x+A_i\vec{x}^\star_m}-\log p_{V^i}\paren{x+A_i\vec{x}^\star_{m^\prime}}}dx
\end{align}
\hrule
\end{figure*}
where $p_{-A_i\vec{x}^\star_m+V^i_1}\paren{x}$ represents the PDF of $-A_i\vec{x}^\star_m+V^i_1$, and $p_{V^i}\paren{x}$ denotes the PDF of $V^i_1$.  Note that $\log p_{V^i}\paren{x+A_i\vec{x}^\star_{m^\prime}}$ can be written as 
\begin{align}\label{Eq: E-E1}
\log p_{V^i}\paren{x+A_i\vec{x}^\star_{m^\prime}}=\log p_{V^i}\paren{x+A_i\vec{x}^\star_{m}-A_i\epsilon^\star_{m,m^\prime}}
\end{align}
where $\epsilon^\star_{m,m^\prime}=\vec{x}^\star_{m}-\vec{x}^\star_{m^\prime}$.  We next use the Taylor expansion Theorem to expand the right hand side of \eqref{Eq: E-E1}. To this end, let $\theta=\left[\theta_1,\cdots,\theta_N\right]^\top$ be an $N$-dimensional vector in $\R^N$. Then, using the Taylor expansion \cite{DK2010} of $\log p_{V^i}\paren{x+A_i\vec{x}^\star_{m}-A_i\theta}$ around $\theta=\vec{0}$, we have 
\begin{align}\label{Eq: TE}
\log p_{V^i}\paren{x+A_i\vec{x}^\star_{m}-A_i\theta}&=\log p_{V^i}\paren{x+A_i\vec{x}^\star_{m}}+\sum_{j=1}^N \theta_j \frac{\partial}{\partial \theta_j} \left.\log p_{V^i}\paren{x+A_i\vec{x}^\star_{m}-A_i\theta}\right|_{\theta=\vec{0}} \nonumber\\
&+\sum_{\abs{\alpha}= 2}\frac{\theta^\alpha}{\alpha !}\partial^\alpha\left.\log p_{V^i}\paren{x+A_i\vec{x}^\star_{m}-A_i\theta}\right|_{\theta=\vec{0}}\nonumber\\
&+\sum_{\abs{\alpha}=3}\frac{\theta^\alpha}{\alpha !}\int_0^13\paren{1-s}^2\partial^\alpha \log p_{V^i}\paren{x+A_i\vec{x}^\star_{m}-sA_i\theta}ds
\end{align}
where $\alpha=\left[\alpha_1,\cdots,\alpha_N\right]^\top$ is an $N$-tuple of positive integers, \emph{i.e.,} $\alpha_i\in\N_0$ for $1\leq i\leq N$, $\abs{\alpha}=\sum_{i}\alpha_i$, $\theta^\alpha=\prod_i\theta_i^{\alpha_i}$, $\alpha !=\prod_i\alpha_i !$,
\begin{align}\label{Eq: FD}
\frac{\partial}{\partial \theta_j} &\left.\log p_{V^i}\paren{x+A_i\vec{x}^\star_{m}-A_i\theta}\right|_{\theta=\vec{0}}=-A_{ij}\frac{\frac{d}{d x}p_{V^i}\paren{x+A_i\vec{x}^\star_{m}}}{p_{V^i}\paren{x+A_i\vec{x}^\star_{m}}}
\end{align}
  and 
\begin{align}\label{Eq: HD}
\partial^\alpha\log p_{V^i}\paren{x+A_i\vec{x}^\star_{m}-A_i\theta}&=\partial^{\alpha_N}\cdots\partial^{\alpha_1}\log p_{V^i}\paren{x+A_i\vec{x}^\star_{m}-A_i\theta}\nonumber\\
&=\paren{-A_{i}}^{\alpha}\frac{d^{\abs{\alpha}}}{d x^{\abs{\alpha}}} \log p_{V^i}\paren{x+A_i\vec{x}^\star_{m}-A_i\theta}
\end{align}

Setting $\theta=\epsilon^\star_{m,m^\prime}$, and substituting \eqref{Eq: TE}-\eqref{Eq: HD}  in \eqref{Eq: KL-Down}, we have \eqref{Eq: KL-Down-1} where $(a)$ follows from the fact that \linebreak $\int_{-\infty}^{\infty}\frac{d}{d x} p_{V^i}\paren{x+A_i\vec{x}^\star_{m}} dx=0$, $\mathcal{I}_i=-\int_{-\infty}^{\infty}p_{V^i}\paren{x}\frac{d^2}{d x^2} \log p_{V^i}\paren{x} dx$ is the Fisher information of the PDF $p_{V^i}\paren{x}$  with respect to a shift parameter, and ${\rm Rem}_i$ is defined in \eqref{Eq: Reminder}.
\begin{figure*}
\begin{align}\label{Eq: KL-Down-1}
&\KLD{p_{-A_i\vec{x}^\star_m+V^i_1}\paren{x}}{p_{-A_i\vec{x}^\star_{m^\prime}+V^i_1}\paren{x}}\nonumber\\
&=A_i\vec{\epsilon}^{\star}_{m,m^\prime}\int_{-\infty}^{\infty}\frac{d}{d x} p_{V^i}\paren{x+A_i\vec{x}^\star_{m}} dx -\frac{1}{2}\paren{ A_i\epsilon^{\star}_{m,m^\prime}}^2\int_{-\infty}^{\infty}p_{V^i}\paren{x+A_i\vec{x}^\star_{m}}\frac{d^2}{d x^2} \log p_{V^i}\paren{x+A_i\vec{x}^\star_{m}} dx\nonumber\\
&-\int_{-\infty}^{\infty}\paren{\sum_{\abs{\alpha}=3}\frac{{\epsilon^\star_{m,m^\prime}}^\alpha}{\alpha !}\paren{-A_{i}}^{\alpha}\int_0^13\paren{1-s}^2\frac{d^3}{d x^{3}} \log p_{V^i}\paren{x+A_i\vec{x}^\star_{m}-sA_i\epsilon^\star_{m,m^\prime}}ds}p_{V^i}\paren{x+A_i\vec{x}^\star_{m}} dx\nonumber\\
&\stackrel{(a)}{=} \frac{1}{2}\paren{ A_i\epsilon^{\star}_{m,m^\prime}}^2\mathcal{I}_i+{\rm Rem}_i
\end{align}
\hrule
\end{figure*}
\begin{figure*}
\begin{align}\label{Eq: Reminder}
{\rm Rem}_i={\int_{-\infty}^{\infty}\paren{\sum_{\abs{\alpha}=3}\frac{{\epsilon^\star_{m,m^\prime}}^\alpha}{\alpha !}\paren{-A_{i}}^{\alpha}\int_0^13\paren{1-s}^2\frac{d^3}{d x^{3}} \log p_{V^i}\paren{x+A_i\vec{x}^\star_{m}-sA_i\epsilon^\star_{m,m^\prime}}ds}p_{V^i}\paren{x+A_i\vec{x}^\star_{m}} dx}
\end{align}
\hrule
\end{figure*}

To complete the proof, we show that ${\rm Rem}_i=\BO{\epsilon^3}$. To this end, we upper bound $\abs{{\rm Rem}_i}$ in \eqref{Eq: Error1} where $(a)$ follows from the assumption 3 in Subsection \ref{Subsec: Non-gaussian}, $(b)$ follows from triangle inequality, $(c)$ follows from the fact that $0\leq s\leq 1$ and $(d)$ follows from the fact that $\norm{\epsilon^\star_{m,m^\prime}}{}=\norm{\paren{\vec{x}^\star_m-\vec{x}^\star_{m^\prime}}}{}\leq 2\sqrt{2}\epsilon$ (see the construction of $\mathcal{S}^\star_{2\epsilon}$ in the proof of Theorem \ref{Theo: Gaussian-bound} for more details).
\begin{figure*}
\begin{align}\label{Eq: Error1}
\abs{\rm Rem_i}&\leq\int_{-\infty}^{\infty}\paren{\sum_{\abs{\alpha}=3}\frac{\abs{{\epsilon^\star_{m,m^\prime}}}^\alpha}{\alpha !}\abs{A_{i}}^{\alpha}\int_0^13\paren{1-s}^2\abs{\frac{d^3}{d x^{3}} \log p_{V^i}\paren{x+A_i\vec{x}^\star_{m}-sA_i\epsilon^\star_{m,m^\prime}}ds}}p_{V^i}\paren{x+A_i\vec{x}^\star_{m}} dx\nonumber\\
&\stackrel{(a)}{\leq} \int_{-\infty}^{\infty}\paren{\sum_{\abs{\alpha}=3}\frac{\abs{{\epsilon^\star_{m,m^\prime}}}^\alpha}{\alpha !}\abs{A_{i}}^{\alpha}\int_0^13\paren{1-s}^2\paren{\beta_1+\beta_2\abs{x+A_i\vec{x}^\star_{m}-sA_i\epsilon^\star_{m,m^\prime}}^{\beta_3}ds}}p_{V^i}\paren{x+A_i\vec{x}^\star_{m}}dx \nonumber\\
&\stackrel{(b)}{\leq}\int_{-\infty}^{\infty}\paren{\sum_{\abs{\alpha}=3}\frac{\abs{{\epsilon^\star_{m,m^\prime}}}^\alpha}{\alpha !}\abs{A_{i}}^{\alpha}\int_0^13\paren{1-s}^2\paren{\beta_1+\beta_2\paren{\abs{x+A_i\vec{x}^\star_{m}}+s\abs{A_i\epsilon^\star_{m,m^\prime}}}^{\beta_3}ds}}p_{V^i}\paren{x+A_i\vec{x}^\star_{m}}dx \nonumber\\
&\stackrel{(c)}{\leq} \int_{-\infty}^{\infty}\paren{\sum_{\abs{\alpha}=3}\frac{\abs{{\epsilon^\star_{m,m^\prime}}}^\alpha}{\alpha !}\abs{A_{i}}^{\alpha}\int_0^13\paren{1-s}^2\paren{\beta_1+\beta_2\paren{\abs{x+A_i\vec{x}^\star_{m}}+\norm{A_i}{}\norm{\epsilon^\star_{m,m^\prime}}{}}^{\beta_3}ds}}p_{V^i}\paren{x+A_i\vec{x}^\star_{m}}dx \nonumber\nonumber\\
&\stackrel{(d)}{\leq}\paren{2\sqrt{2}\epsilon}^3 \sum_{\abs{\alpha}=3}\frac{\abs{A_{i}}^{\alpha}}{\alpha !}\!\paren{\int_0^13\paren{1-s}^2ds}\!\!\paren{\beta_1+\int_{-\infty}^{\infty}\beta_2\paren{\abs{x+A_i\vec{x}^\star_{m}}+2\sqrt{2}\norm{A_i}{}\epsilon}^{\beta_3}\!\!p_{V^i}\paren{x+A_i\vec{x}^\star_{m}}dx} 
\end{align}
\hrule
\end{figure*}
Note that the second integral in the right hand side of \eqref{Eq: Error1} can be upper bounded as \eqref{Eq: Error2}.
\begin{figure*}
\begin{align}\label{Eq: Error2}
\int_{-\infty}^{\infty}\beta_2\paren{\abs{x+A_i\vec{x}^\star_{m}}+2\sqrt{2}\norm{A_i}{}\epsilon}^{\beta_3}p_{V^i}\paren{x+A_i\vec{x}^\star_{m}}dx&=\int_{-\infty}^{\infty}\beta_2\paren{\abs{x}+2\sqrt{2}\norm{A_i}{}\epsilon}^{\beta_3}p_{V^i}\paren{x}dx\nonumber\\
&\leq \beta_2\sum_{j=1}^\infty\paren{j+2\sqrt{2}\norm{A_i}{}\epsilon}^{\beta_3}\PR{\abs{V^i_1}\geq j-1}\nonumber\\
&= \beta_2\sum_{j=1}^\infty\paren{1+2\sqrt{2}\frac{\norm{A_i}{}\epsilon}{j}}^{\beta_3}j^{\beta_3}\PR{\abs{V^i_1}\geq j-1}\nonumber\\
&\leq \beta_2\paren{1+2\sqrt{2}\norm{A_i}{}\epsilon}^{\beta_3}\sum_{j=1}^\infty j^{\beta_3}\PR{\abs{V^i_1}\geq j-1}
\end{align}
\hrule
\end{figure*}
 It is straightforward to show that the series $\sum_{j=1}^\infty j^{\beta_3}\PR{\abs{V^i_1}\geq j-1}$ in  \eqref{Eq: Error1} is bounded since the PDF of $V^i_1$, \emph{i.e.,} $p_{V^i}\paren{x}$, satisfies $\lim_{x\rightarrow\infty}x^{\paren{\beta_3+1+r}}\PR{\abs{V^i_1}\geq x }=0$ for some $r>0$. Thus, we have ${\rm Rem}_i=\BO{\epsilon^3}$ which completes the proof.
\end{IEEEproof}

 Using \eqref{Eq: KL-2} and \eqref{Eq: KL-Down-1}, $\MI{\vec{x}^\star_M}{-A\vec{x}^\star_M+V_1}$ can be upper bounded as \eqref{Eq: ML-F} where $(a)$ follows from the facts that $\sum_i\norm{A_i}{}^2\leq N\norm{A}{}$ and $\max_{\vec{x}^\star_m,\vec{x}^\star_{m^\prime}\in\mathcal{S}^\star_{2\epsilon}}\norm{\paren{\vec{x}^\star_m-\vec{x}^\star_{m^\prime}}}{}=2\sqrt{2}\epsilon$, and $(b)$ follows from the fact that $\norm{A}{}=\gamma$.
\begin{figure*}
\begin{align}\label{Eq: ML-F}
\MI{\vec{x}^\star_M}{-A\vec{x}^\star_M+V_1}&\leq  \max_{\vec{x}^\star_m,\vec{x}^\star_{m^\prime}\in\mathcal{S}^\star_{2\epsilon}}\sum_{i=1}^N\frac{1}{2}\mathcal{I}_i\paren{ A_i\paren{\vec{x}^\star_{m}-\vec{x}^\star_{m^\prime}}}^2+{\rm Rem}_i\nonumber\\
&\leq  \max_{\vec{x}^\star_m,\vec{x}^\star_{m^\prime}\in\mathcal{S}^\star_{2\epsilon}}\sum_{i=1}^N\frac{1}{2}\mathcal{I}_i\paren{ A_i\paren{\vec{x}^\star_{m}-\vec{x}^\star_{m^\prime}}}^2+\max_{\vec{x}^\star_m,\vec{x}^\star_{m^\prime}\in\mathcal{S}^\star_{2\epsilon}}\sum_{i=1}^N{\rm Rem}_i\nonumber\\
&\leq  \max_{\vec{x}^\star_m,\vec{x}^\star_{m^\prime}\in\mathcal{S}^\star_{2\epsilon}}\sum_{i=1}^N\frac{1}{2}\mathcal{I}_i\norm{\vec{x}^\star_{m}-\vec{x}^\star_{m^\prime}}{}^2\norm{A_i}{}^2+\max_{\vec{x}^\star_m,\vec{x}^\star_{m^\prime}\in\mathcal{S}^\star_{2\epsilon}}\sum_{i=1}^N{\rm Rem}_i\nonumber\\
&\stackrel{(a)}{\leq}  4\epsilon^2N\norm{A}{}^2\max_i\mathcal{I}_i+\max_{\vec{x}^\star_m,\vec{x}^\star_{m^\prime}\in\mathcal{S}^\star_{2\epsilon}}\sum_{i=1}^N{\rm Rem}_i\nonumber\\
&\stackrel{(b)}{=}  4N\epsilon^2\gamma^2\max_i\mathcal{I}_i+\max_{\vec{x}^\star_m,\vec{x}^\star_{m^\prime}\in\mathcal{S}^\star_{2\epsilon}}\sum_{i=1}^N{\rm Rem}_i
\end{align}
\hrule
\end{figure*} 
Since we have ${\rm Rem}_i=\BO{\epsilon^3}$ (see the proof of Lemma \ref{Lem: KL-expansion}), \eqref{Eq: ML-F}  implies 
\begin{align}\label{Eq: MI-NG}
\MI{\vec{x}^\star_M}{-A\vec{x}^\star_M+V_1}\leq 4N\epsilon^2\gamma^2\max_i\mathcal{I}_i+\BO{\epsilon^3}
\end{align}

Combining \eqref{Eq: E-E0} and \eqref{Eq: MI-NG}, we have 
\begin{eqnarray}
T\geq \frac{2\paren{1-\delta}-1}{4N\epsilon^2\gamma^2\max_i\mathcal{I}_i+\BO{\epsilon^3}}\nonumber
\end{eqnarray}
which completes the proof.

%%%%%%%%%%%%%%%%%%%%%%%%%%%%%%%%%%%%%%%%%%%%%%%%%%%%%%%%%%%%%%%%%%%%%%%%%%%%%%%%%%%%%%%%%%%%%%
\subsection{Proof of Theorem \ref{Theo: MI-bound}}\label{App:  MI-bound}
To establish this result, we first construct a finite subset of $\mathcal{F}_\gamma$. To this end,  let $\mathcal{S}_{2\epsilon}$ denote an \emph{arbitrary} $2\epsilon$-distinguishable subset of $\mathcal{S}$. Note that the set $\mathcal{S}_{2\epsilon}$ is not necessarily a maximal size $2\epsilon$-distinguishable subset of $\mathcal{S}$.  For each $\vec{x}^\star_m\in\mathcal{S}_{2\epsilon}$ ($m=1,\cdots,\abs{\mathcal{S}_{2\epsilon}}$), we construct a utility function vector $U_m\paren{\vec{x}}$ such that $\vec{x}^\star_m$ is the NE of the non-cooperative game with $N$ players, utility function vector $U_m\paren{\vec{x}}$ and the action space $\mathcal{S}$.

The utility function vector $U_m\paren{\vec{x}}$ ($m=1,\cdots,\abs{\mathcal{S}_{2\epsilon}}$) is constructed as follows. Let $A=\left[a_{ij}\right]_{i,j}$ be an $N$-by-$N$, symetric, negative definite matrix with $\norm{A}{}\geq \gamma$. The utility function vector $U_m\paren{\vec{x}}$ is defined as $U_m\paren{\vec{x}}=\left[u_{m,i}\paren{x^i,\vec{x}^{-i}}\right]_i$ where $u_{m,i}\paren{x^i,\vec{x}^{-i}}=\frac{a_{ii}}{2}\paren{x^i}^2+x^i\paren{-A_i\vec{x}^\star_m+\sum_{j\neq i} a_{ij}x^j}$ where $A_i$ is the $i$th row of $A$. Let $\mathcal{F}^\prime$ be the finite set of utility function vectors defined as 
\begin{eqnarray}
\mathcal{F}^\prime_\gamma=\left\{U_m\paren{\cdot}\in\mathcal{F}, m=1,\cdots,\abs{\mathcal{S}_{2\epsilon}}\right\}
\end{eqnarray}
Clearly, we have $\abs{\mathcal{F}^\prime}=\abs{\mathcal{S}_{2\epsilon}}$

 It is straight forward to verify that $U_{m}\paren{\vec{x}}$ belongs to $\mathcal{F}_{\gamma}$ which implies that ${\mathcal{F}}_\gamma^\prime$ is a subset of $\mathcal{F}_\gamma$. We refer to the $N$-player non-cooperative games with the utility functions in $\mathcal{F}^\prime_\gamma$ and the action space $\mathcal{S}$ as ${\mathcal{G}}^\prime_\gamma=\langle \mathcal{N}, \mathcal{S},{\mathcal{F}}^\prime_\gamma\rangle$. Similar to the proof of Theorem 1, we make the technical assumption that each game in ${\mathcal{G}}_\gamma^\prime$ admits a unique Nash equilibrium (NE). 

Now, for a given $\epsilon$ and $\delta$,  consider any algorithm $\mathcal{A}$ for which after $T$ time-steps, we have 
\begin{align}
\!\!\!\!\sup_{U\paren{\cdot}\in \mathcal{F}_\gamma}\!\!\inf_i\PRP{\norm{\vec{x}_{{\rm NE}_i, U\paren{\cdot}}\!\!-\!\!\mathcal{A}_{T+1}\!\!\paren{{X}_{1:T},{\hat{Y}_{1:T}},\hat{Z}_{1:T}}}{}\geq \epsilon}\leq \delta.\nonumber
\end{align}
 Since ${\mathcal{F}}^\prime_\gamma$ is a subset of $\mathcal{F}_\gamma$ and each game in ${\mathcal{G}}^\prime_\gamma$ admits a unique NE, we have 
\begin{align}
\sup_{m=1,\cdots,\abs{{\mathcal{S}}_{2\epsilon}}}\PRP{\norm{\vec{x}^\star_{m}-\mathcal{A}_{T+1}\paren{{X}_{1:T},{\hat{Y}_{1:T}},\hat{Z}_{1:T}}}{}\geq \epsilon}\leq \delta.\nonumber
\end{align}
Using Lemma \ref{Lem: HT} in Subsection \ref{Subsec: HT} and Fano inequality, we have 
\begin{align}\label{Eq: D-E1}
\delta\geq 1-\frac{1+\MI{M}{\hat{M}}}{\log\abs{{\mathcal{S}}_{2\epsilon}}}
\end{align}
Combing \eqref{Eq: C-E2} and \eqref{Eq: D-E1}, we have 
\begin{eqnarray}
T\geq \frac{\paren{1-\delta}\log\abs{{\mathcal{S}}_{2\epsilon}}-1}{\MI{\vec{x}^\star_M}{-A\vec{x}^\star_M+\hat{W}_A}}\nonumber
\end{eqnarray}
where $\vec{x}^\star_M$ is a random vector taking value in ${\mathcal{S}}_{2\epsilon}$ with uniform distribution and $\hat{W}_A=\left[\hat{W}^i_1\right]_i$ with \linebreak $\hat{W}^i_1=\sum_{j\in\mathcal{N}_{{\rm usn}_{\pi\paren{i}}}}\paren{a_{ij}W^{j}_{1,{\rm usn}_{\pi\paren{i}}}}+V^i_{1}$. Optimizing over the choice of the matrix $A$ and the $2\epsilon$-distinguishable set ${\mathcal{S}}_{2\epsilon}$, we have 
\begin{eqnarray}
T\geq\sup_{A:\norm{A}{}\geq 2,{\mathcal{S}}_{2\epsilon}\in\mathcal{S}} \frac{\paren{1-\delta}\log\abs{{\mathcal{S}}_{2\epsilon}}-1}{\MI{\vec{x}^\star_M}{-A\vec{x}^\star_M+\hat{W}_A}}\nonumber
\end{eqnarray}
which completes the proof.

\section{Conclusion}\label{Sec: Conc}
\textcolor{black}{In this paper, we  studied the complexity of solving two game classes  in a distributed setting in which  players obtain the required information for updating their actions by communicating with a set of system nodes over noisy communication channels. We first considered the game class $\mathcal{G}$ which is comprised of all $N$-player non-cooperative games with a continuous action space such that any game in $\mathcal{G}$ admits at least a Nash equilibrium. We obtained a lower bound on the complexity of solving the game class $\mathcal{G}$ to an $\epsilon$ accuracy which depends on the Kolmogorov $2\epsilon$-capacity of the constraint set and the total capacity of the communication channels which convey utility-related information to players. We also studied the complexity of solving a subclass of $\mathcal{G}$ under both Gaussian and non-Gaussian noise models.}
\bibliographystyle{IEEEtran}
\bibliography{Game_Complexity} 

\end{document}